\documentclass[pre,aps,superscriptaddress,tightenlines,twocolumn,showpacs,floatfix,longbibliography,nofootinbib]{revtex4-1}
\usepackage{latexsym,amsmath,amsfonts,amssymb,bm,empheq}
\usepackage{amsmath,amssymb}
\usepackage{graphicx}
\usepackage{psfrag}
\usepackage{color}
\usepackage{dcolumn}
\include{ckpzdisorder.bib}
\usepackage{bm}
\usepackage[normalem]{ulem}

\def\beq{\begin{equation}}
\def\eeq{\end{equation}}
\def\bea{\begin{eqnarray}}
\def\eea{\end{eqnarray}}
\usepackage{hyperref}
\hypersetup{
 colorlinks=true,
 citecolor=blue,
}

\begin{document}

\title{
Disorders can induce continuously varying universal scaling in driven systems}
\author{Astik Haldar}\email{astik.haldar@gmail.com, astik.haldar@saha.ac.in}
\affiliation{Theory Division, Saha Institute of
Nuclear Physics, HBNI, Kolkata 700064, India}
\author{Abhik Basu}\email{abhik.123@gmail.com, abhik.basu@saha.ac.in}
\affiliation{Theory Division, Saha Institute of
Nuclear Physics, HBNI, Kolkata 700064, India}

\begin{abstract}
We elucidate the nature of universal scaling in a class of quenched disordered driven models. In particular, we explore the intriguing possibility of whether coupling with quenched disorders can lead to {\em continuously varying} universality classes. We examine this question in the context of the Kardar-Parisi-Zhang (KPZ) equation, with and without a conservation law, coupled with quenched disorders  having distributions with pertinent structures. We show that when the disorder is relevant in the renormalisation group sense, the scaling exponents can depend continuously on a dimensionless parameter that defines the disorder distribution. This result is generic, and holds for quenched disorders with or without spatially long ranged correlations, as long as the disorder remains a ``relevant perturbation'' on the pure system in the renormalisation group sense and a dimensionless parameter naturally exists in its distribution. We speculate on its implications for generic driven systems with quenched disorders, and compare and contrast with the scaling displayed in the presence of annealed disorders.
\end{abstract}

\maketitle

\section{Introduction}

Classification of the physics of nonequilibrium systems at long time and
length scales into universality classes remains a theoretically challenging task.
The standard universality classes in critical dynamics are quite
robust to detailed-balance violating perturbations~\cite{tauber-santos}. 
In contrast, genuine nonequilibrium dynamic phenomena, having
 non-Gibbsian steady states, are found to be rather sensitive to all kinds of perturbations.
Notable examples include driven diffusive systems~\cite{driven}. For instance, for the Kardar-Parisi-Zhang (KPZ)
equation, anisotropic perturbations are found to be relevant in all dimensions $d\geq 2$~\cite{tauber-frey}.

Equilibrium systems either in the vicinity of critical points or in broken symmetry phases for systems with continuous symmetries show universal scaling that depend only on the spatial dimension $d$ and the symmetry
of the order parameter (e.g., Ising, XY, nematic etc)~\cite{fisher-lectures}, but are independent of the material parameters that define the model. Prominent exceptions are the two-dimensional (2D) XY model and its related models, where the long wavelength universal properties are controlled not by a fixed point but a fixed line, and
as a result, the scaling exponents of the relevant correlation functions exhibit a continuous dependence on the stiffness
parameters~\cite{chaikin}. Furthermore, equilibrium dynamics close to critical points also show dynamic universality though
the dynamic scaling exponents, which characterise the time-dependence of the unequal-time
correlation functions, and now may also depend upon the presence or absence of conservation
laws and the non-dissipative terms in the dynamical equations~\cite{halperin-review}. Genuinely nonequilibrium systems can also show the surprising features of continuously varying nonequilibrium universal properties, which are paramatrised by dimensionless parameters that naturally appear in the formulations of the dynamical models. Well-known examples include stochastically forced fully developed three-dimensional magnetohydrodynamic turbulence~\cite{abmhd} and stochastically coupled Burgers equations~\cite{ab-burgers1,ab-burgers2}. In all these examples, the relevant dimensionless parameters appear in the definition of the noise variances. More recently, it has been suggested that a nearly phase-ordered collection of diffusively mobile, active XY spins on a substrate can be stable, and the relevant scaling exponents that describe the phase and density correlation functions can vary continuously with certain anharmonic coupling constants that appear in the hydrodynamic equations of motion~\cite{diff-xy-short,diff-xy-long}.

Disorder is known to affect the large-scale, long time universal properties of condensed matter systems. Depending upon the time-scales, disorders are classified into two classes - quenched and annealed. Quenched disorders are frozen in time and do not thermalise even in equilibrium systems. In contrast, annealed disorders time-evolve and thermalise in the long time limit in equilibrium systems. Effects of quenched disorder on the universal properties of equilibrium systems are well-documented. For example, quenched disorders that locally affect the critical temperature lead to   new universality classes different from the corresponding pure model; see Refs.~\cite{quench-equi,halperin-disorder,quench-on1}. Unlike quenched disorders, annealed disordered equilibrium systems may be viewed as pure systems supplemented by additional thermal degrees of freedom. This may result into different static and dynamic universality, depending upon the systems under consideration~\cite{chaikin,tirtha-membrane1,tirtha2}. 

Effects of quenched disorders on nonequilibrium systems can be strikingly more complex. For instance, the random Gaussian-distributed quenched columnar disordered Kardar-Parisi-Zhang (KPZ) equation, which in one dimension (1D) reduces to the periodic totally asymmetric simple exclusion process (TASEP) with short-ranged Gaussian-distributed quenched disordered hopping rates is affected by the disorder only in certain special 
limits of the model parameters (in 1D which is equivalent to the half-filled limit in the TASEP language), leading to  new universal scaling behaviour in all dimensions $d$. Else, pure KPZ universality holds~\cite{astik-prr,astik-pre}. In a recent study, it has been shown that the conserved KPZ (CKPZ) equation, when coupled with a particular choice of quenched disorder can modify the scaling behaviour of the pure CKPZ equation in some cases~\cite{sudip-ckpz}. Nonetheless, the question of disorder-induced continuous universality in driven models remains largely unexplored till the date.

In this article, we explore the possibility of  disorder-induced continuous universality in driven models. In the absence of any general theoretical framework for nonequilibrium systems, it is useful to study simple models where such questions can be explored systematically by using analytically tractable calculations. Insights drawn from such studies should be useful to enhance general understanding of scaling in nonequilibrium setting. To that end, in this work, we have studied two models - one with a conservation law, the conserved KPZ (CKPZ) equation and the other without any conservation, the KPZ equation, both coupled with ``orientational'' disorders, which couples with the local gradient of the height field. By carefully choosing the disorder distributions, we show that the resulting universality classes, when they depend upon the quenched disorders, can vary continuously with a dimensionless parameter that characterises the disorder distributions. In addition, we briefly study a version of annealed disordered CKPZ equation, and show that similar continuously varying scaling exponents can be found under certain circumstances. The remainder of the paper is organised as follows. In Section~\ref{kpz-ckpz}, we review the pure KPZ and CKPZ universality classes. Next in Section~\ref{model}, we introduce the disordered KPZ and CKPZ equations that we use here. In Section~\ref{rg}, we elucidate the scaling properties of the quenched disordered KPZ and CKPZ equations. In Section~\ref{summ}, we summarise our results. We have used one-loop dynamic renormalisation group (RG) calculations for our work. We discuss some of the technical details in Appendix for interested readers. We  briefly analyse the case of annealed disordered CKPZ equation in Appendix.



 \section{Equations for fluctuating surfaces}\label{kpz-ckpz}
 
 We consider a fluctuating surface without overhangs that may be moving (i.e., growing) on average, or not moving. Microscopically, these
growth processes are described by local dynamics or local update rules, and are generic examples of nonequilibrium driven systems. An interface in $d+1$-dimensional hyperspace is characterised by variations of height $h({\bf x},t)$, where  ${\bf x}(x_1,x_2,....,x_d)$ is a position vector in the $d$-dimensional hypersurface. We consider fluctuating surfaces having dynamics unaffected by their absolute heights with respect to specific base planes. In other words, the dynamics is invariant under constant shifts of $h$.  The height fluctuations display dynamic scaling~\cite{stanley}, and the associated time-dependent correlation function in the steady states has a scaling form
%
\begin{align}
 C(r,t-t')\equiv\langle [ h({\bf x},t) - h({\bf x'},t')]^2 \rangle=r^{2\chi_h}\varphi\left(\frac{r^z}{t-t'}\right) \label{exponent}
\end{align}
in the long wavelength limit,
where $r\equiv |\bf x-x'|$; $\chi_h$ and $z$ are  {\em roughness exponent} and {\em dynamical exponent} respectively, which classify the universality class;  $\varphi$ is a dimensionless scaling function. While the values of the scaling exponents are  independent of the model parameters, they of course vary from one universality class to another. Two such well-known universality classes are those associated with the KPZ and CKPZ equations, which we review briefly below.


 \subsection{KPZ universality class}\label{pure-kpz}
 The  KPZ equation, originally proposed as a surface growth model~\cite{kpz}, is the paradigmatic nonequilibrium model that shows nonequilibrium phase transitions at $d>2$. It is given by
 \begin{align}
  \frac{\partial h}{\partial t}=\nu_1 \nabla^2 h - \frac{\lambda_1}{2}(\boldsymbol\nabla h)^2 + \xi_h.\label{kpz}
 \end{align}
 Here, $\nu_1>0$ is a diffusivity, and $\lambda_1$ is a nonlinear coefficient. Furthermore, $\xi_h$ is noise with Gaussian distribution and zero mean, that is added to describe the inherent stochasticity of the dynamics. It is a white noise, since $h$ in (\ref{kpz}) obeys a non-conserved dynamics.  
   Stochastic noise $\xi_h$ has a variance
 \begin{align}
  \langle \xi_h({\bf x},t) \xi_h({\bf x'},t') \rangle= 2D_1\delta^d({\bf x-x'})\delta(t-t'), \label{xi}
 \end{align}
  where $D_1>0$ is the noise strength. { In addition to its application as a surface growth model, (\ref{kpz}) serves as the active or nonequilibrium hydrodynamic model for a smoothly varying phase, when there is no other hydrodynamic variable present~\cite{john-prx}, and an {\em active fluid membrane} without momentum conservation~\cite{active-mem}} .
  
  
  Equation~(\ref{kpz}) is invariant under the transformation ${\bf x}\rightarrow {\bf x}-\lambda_1{\bf a}t,\, t\rightarrow t,\, h\rightarrow h+{\bf a\cdot x}+\frac{\lambda_1}{2}|{\bf a}|^2t$, known as the tilt invariance~\cite{stanley}. This in turn gives an exact exponent relation $\chi_h+z=2$~\cite{stanley,erwin-two-loop}.  In 1D  the scaling exponents are found exactly  as a consequence of the tilt invariance (that holds at all dimensions) and the Fluctuation-Dissipation-Theorem that holds only at 1D~\cite{stanley}. This gives dynamic exponent $z=1/2$ and roughening exponent $\chi_h=3/2$ {\em exactly}, corresponding to a 1D rough surface. In 2D, which is the {\em lower critical dimension} of this model, there is only a ``rough'' phase which is perturbatively inaccessible. Furthermore, the KPZ equation has smooth phase above 2D for low enough noise, the scaling property of which is identical to the  linear Edward-Wilkinson (EW) equation~\cite{stanley} with $z=2, \chi_h=\frac{2-d}{2}$. As the noise strength is increased, the KPZ equation undergoes a phase transition from a smooth to a perturbatively inaccessible rough phase in dimension $d>2$.

 \subsection{CKPZ universality class}\label{pure-ckpz}
 
 The CKPZ equation, which is essentially the conserved analogue of the KPZ equation, forms a universality class distinct from the KPZ equation. This of course is not surprising, since in nonequilibrium systems, the presence or absence of conservation laws not only affect the dynamic scaling exponents, they can in principle affect the static exponents (e.g., the roughness exponent) as well, in contrast to equilibrium systems. In the CKPZ equation, the height field $h({\bf x, t})$ of an interface of a volume conserving system, a single-valued function measured with respect to an arbitrary base plane, follows a generic conservation laws:
  \begin{align}
 &\partial_t h= -\boldsymbol\nabla \cdot {\bf J},
\end{align}
where ${\bf J}$, the  current, has the following form~\cite{ckpz}
\begin{align}
 {\bf J}= \boldsymbol\nabla\left[\nu_2 \nabla^2 h -\frac{\lambda_2}{2}(\boldsymbol\nabla h)^2 \right].\label{jeq} 
\end{align}
We note that current $\bf J$ in (\ref{jeq}) is constructed in such a way that ${\bf J}({\bf k=0},t)=0$, where $\bf k$ is a wavevector. With this, the CKPZ equation takes the form~\cite{ckpz}  
\begin{align}
   \frac{\partial h}{\partial t}=-\nabla^2\left[\nu_2 \nabla^2 h - \frac{\lambda_2}{2}(\boldsymbol\nabla h)^2\right] + \eta_h.\label{ckpz}
 \end{align}
 Here,   $\eta_h$ is a conserved noise that models stochastic nature of dynamics.  It is assumed to be zero-mean and Gaussian distributed with a variance
 \begin{align}
  \langle \eta_h({\bf x},t) \eta_h({\bf x'},t') \rangle= 2D_2(-\nabla^2)\delta^d({\bf x-x'})\delta(t-t'). \label{eta}
 \end{align} 
 In the linear limit, i.e., with $\lambda_2=0$, Eq.~(\ref{ckpz}) reduces to the Mullins-Herring (MH) equation for linear MBE processes~\cite{mull,herr}. 
Equation~(\ref{ckpz}) is not tilt invariant, and hence, it has no exact exponent relation in contrast to the KPZ equation~\cite{janssen}.  Results from one-loop RG study shows that $d=2$ is the upper critical dimension, and below 2D, in an  $\epsilon= 2-d$ expansion  one finds the scaling exponents as $z=4-\frac{\epsilon}{3},\, \chi_h=\frac{\epsilon}{3}$. This corresponds to a  rough phase. At 2D the interface is {\em logarithmically} rough and above 2D, the surface is smooth, with long wavelength scaling properties statistically identical to those obtained from the linear MH equation. Unlike the KPZ equatiom, the CKPZ equation does not admit a  smooth-to-rough transition for strong coupling~\footnote{A generalised CKPZ equation has recently been proposed that contains an additional nonlinear term which is as relevant (in a RG sense) as the existing nonlinear term in (\ref{ckpz}). This admits a roughening transition; see  F. Caballero et al, Strong Coupling in Conserved Surface Roughening: A New Universality Class?, Phys. Rev. Lett. {\bf 121}, 020601 (2018). We do not discuss that here.}.

 
\section{Disordered KPZ and CKPZ equations}\label{model}

We now minimally couple the KPZ Eq.~(\ref{kpz}) and CKPZ Eq.~(\ref{ckpz}) equations  with ``orientational'' quenched disorder to address the question on universality we raised above. We call it an orientational disorder, since it, a random quenched disordered vector field ${\bf V}({\bf x})$, whose statistics is given below, couples with the fluctuation in the local orientation of the height field given by ${\boldsymbol\nabla} h$. Further,  $V_i$ can in general have both irrotational and solenoidal parts. Quenched vector $V_i$ is assumed to be zero-mean, Gaussian-distributed with a variance
\begin{align}
 \langle V_i({\bf x}) V_j({\bf x'}) \rangle= \left[2D_T P_{ij}+2D_L Q_{ij}\right]|{\bf x-x'}|^{-\alpha}. \label{veq}
\end{align}
Here $P_{ij}$  and  $Q_{ij}$ are the transverse (solenoidal) and longitudinal (irrotational) projection operators respectively. 
In the Fourier space, these are given by
\begin{align}
 P_{ij}({\bf k})=\delta_{ij}-\frac{k_i k_j}{k^2} ,\, Q_{ij}({\bf k})=\frac{k_i k_j}{k^2}.\label{component} 
\end{align}
Thus, $P_{ij}({\bf k})$ and $Q_{ij}({\bf k})$, respectively,  project any vector they operate on to directions normal and parallel to $\bf k$.  Evidently at 1D, $P_{ij}\equiv 0$ identically. Noise strengths $D_T$ and $D_L$ are positive definite. The exponent $\alpha$ paramatrises the  measure of how {\em long range} or how spatially correlated the quenched disorder is; we take $0\le\alpha<d$ { (in this range the short distance cutoff implied in the Fourier transforms can be smoothly taken to zero)}.
  In the Fourier space the disorder correlation (\ref{veq}) takes the form:
 \begin{align}
  &\langle V_i({\bf k},\omega) V_j({\bf k'},\omega') \rangle= \left[2D_T P_{ij}+2D_L Q_{ij}\right] k^{-\mu}\nonumber\\
 &~~~~~~~~~~~~~~~~~~~~\times \delta^d({\bf k+k'}) \delta(\omega')\delta(\omega),\label{v-var}
 \end{align}
 where $\mu\equiv d-\alpha >0$ for spatially long-ranged correlated disorders. {\ In the  more familiar short-ranged disorder case 
\begin{equation}
 \langle V_i({\bf x}) V_j({\bf x'}) \rangle= \left[2D_T P_{ij}+2D_L Q_{ij}\right]\delta^d(\bf{x-x'})\label{v-short}
\end{equation}
in real space. In the Fourier space, this gives (\ref{v-var}) with $\mu=0$.}
Clearly, when $D_L=D_T$, the rhs of (\ref{v-short}) is proportional to $\delta_{ij}$~\cite{sudip-ckpz}. Further, in the extreme limits when $D_T=0$, the quenched vector field ${\bf V}$ is irrotational, whereas for $D_L=0$, it is solenoidal. 

In the next Sections, we study the effects of the coupling of $V_i$ with the KPZ and CKPZ equations, and explore their universal scaling properties, in particular, the dependence on the dimensionless ratio $\gamma\equiv D_T/D_L$.

 \subsection{KPZ equation with quenched disorder}\label{disordered-kpz}
 
 { Given the status of the KPZ equation as a paradigmatic nonequilibrium model that shows phase transitions, it is of great theoretical interest to study whether or not quenched disorder can modify the universality classes of the pure system. }
 The quenched disordered version of the KPZ equation - the KPZ equation with minimally coupled orientational quenched disorder that we use is
 \begin{align}
 \frac{\partial h}{\partial t}=\nu_1 \nabla^2 h - \frac{\lambda_1}{2}(\boldsymbol\nabla h)^2 +\kappa_1 (\boldsymbol{V}\cdot \boldsymbol\nabla h)+ \xi_h.\label{heq-kpz}
 \end{align}
  The disorder-dependent nonlinear  term with coefficient $\kappa_1$ is the  leading order nonlinear term that respects the invariance under a constant shift of $h$ (see Ref.~\cite{anton-1} for a similar coupling). It breaks the tilt invariance of the pure KPZ equation. As a result, Eq.~(\ref{heq-kpz}) lacks any exact exponent identity, in direct contrast with the pure KPZ equation. Further, the sign of $\kappa_1$ is arbitrary. Equation~(\ref{heq-kpz}) reduces to the pure KPZ equation (\ref{kpz}) for $\kappa_1=0$.   The additive annealed noise $\xi_h$ is assumed to be zero-mean, Gaussian-distributed with a variance given by (\ref{xi}). { Just as the pure KPZ equation (\ref{kpz}) is the nonequilibrium hydrodynamic model of a phase in the absence of any other hydrodynamic variables, (\ref{heq-kpz}) is the nonequilibrium hydrodynamic model of a smoothly varying phase field in the presence of orientational quenched disorder in the absence of any other hydrodynamic variables, or of an active fluid membrane with quenched disorder with broken tilt invariance and without momentum conservation. Thus this study should provide insight to the role of quenched disorders in these systems.}

\subsection{CKPZ equation with quenched disorder}\label{disordered-ckpz} 


To study the CKPZ equation minimally coupled with orientational quenched disorder, we use the disordered version of the CKPZ equation proposed and studied in Ref.~\cite{sudip-ckpz}. The form of the current $\bf J$ corresponding to this disordered CKPZ equation reads
\begin{align}
 {\bf J}= \boldsymbol\nabla\left[\nu_2 \nabla^2 h -\frac{\lambda_2}{2}(\boldsymbol\nabla h)^2 -\kappa_2 (\boldsymbol{V}\cdot \boldsymbol\nabla h)\right].\label{jeq-dis} 
\end{align}
The coefficient $\kappa_2$ is the coupling constant of the disorder-dependent  leading order nonlinear term that respects the invariance under a constant shift of $h$; $\kappa_2$ can take any sign. The choice of disorder coupling that current ${\bf J}({\bf k}\rightarrow 0, t) \rightarrow 0$ at the thermodynamic limit. With $\kappa_2=0$, the model reduces to the pure CKPZ equation (\ref{ckpz}). With (\ref{jeq-dis}), the disordered CKPZ equation reads 
\begin{align}
 &\partial_t h=\nabla^2\left[-\nu_2 \nabla^2 h +\frac{\lambda_2}{2}(\boldsymbol\nabla h)^2 +\kappa_2 (\boldsymbol{V}\cdot \boldsymbol\nabla h)\right] +\eta_h. \label{heq-ckpz}
\end{align} 
  The noise $\eta_h$ satisfy the same as Eq. (\ref{eta}).

 In the next section we analyse the scaling properties of these disordered models.


\section{Universal scaling in the disordered models}\label{rg}

The universality classes of the pure KPZ and CKPZ equations are well-established. We now set out to find whether the quenched disorder is a relevant perturbation on these universality classes, and if so, what the new universality classes are. 
The nonlinear terms present in (\ref{heq-kpz}) and (\ref{heq-ckpz}) preclude any exact analysis of the problem, and necessitate use of perturbative approaches.  The na\"ive perturbation theory produces diverging corrections to the model parameters in the long wavelength limit. We use here the dynamic RG framework to systematically handle these diverging corrections in the long wavelength limit. We outline this method below. It is convenient to express the stochastically driven equations (\ref{heq-kpz}) and (\ref{heq-ckpz}) as path integrals over configurations of $h({\bf r},t)$ and its dynamic conjugate field $\hat h({\bf r},t)$~\cite{bausch,uwe-book}, subject to the distribution of the quenched disorders as specified above. The momentum shell RG procedure consists of integrating over the short wavelength 
Fourier modes
of $h({\bf r},t)$, $\hat h({\bf r},t)$ and $V_i({\bf r})$, followed by  rescaling of lengths and times~\cite{uwe-book}. In particular, we  follow the usual convention of initially restricting the wavevectors  
to be within a bounded spherical Brillouin zone: $|{\bf k}|<\Lambda$. However, the precise value of the upper cutoff $\Lambda$ has no effect on our final  results. The fields $h({\bf r},t)$, $\hat h({\bf r},t)$ and $V_i({\bf r})$
are separated into the high and low wave vector parts
$h({\bf r},t)=h^<({\bf r},t)+h^>({\bf r},t)$, $\hat h({\bf r},t)=\hat h^<({\bf r},t) + \hat h^>({\bf r},t)$ and $V_i({\bf r})=V_i^<({\bf r}) + V_i^>({\bf r})$, where $h^>({\bf r},t)$, $\hat h^>({\bf r},t)$ and $V_i({\bf r})^>$ have support in the large wave vector  (short wavelength) 
range $\Lambda
e^{-l}<|{\bf k}|<\Lambda$, while $h^<({\bf r},t)$, $\hat h^<({\bf r},t)$ and $V_i^<({\bf r})$ have support in the small 
wave vector (long wavelength) range $|{\bf k}|<e^{-l}\Lambda$; $b\equiv e^{l}>1$.
We then integrate out $h^>({\bf r},t)$, $\hat h^>({\bf r},t)$ and $V_i^>({\bf r})$ perturbatively 
in 
 the anhamornic couplings, which can only be done perturbatively; as usual, this resulting perturbation theory of $h^<({\bf r},t)$, $\hat h^<({\bf r},t)$ and $V_i^<({\bf r})$
can be represented by Feynman graphs, with the order of perturbation theory 
reflected by the number of loops in the graphs we consider; see, e.g., Refs.~\cite{astik-prr,astik-pre}. 
This procedure allows us to calculate the RG flow equations which give the stable fixed points of the disordered KPZ and CKPZ equations, which in turn give the associated scaling exponents. In the next Sections, we analyse the RG flow equations and elucidate the scaling exponents for the quenched disordered KPZ and CKPZ equations separately. Interested readers will find
the relevant one-loop Feynman diagrams and other details of the intermediate steps in Appendix~\ref{DRG}.

\subsection{Quenched disordered KPZ equation}\label{kpz-rg}
The dynamic exponent  and roughness exponent in the linear limit of (\ref{heq-kpz}) 
are $z=2$ and $\chi_h=\frac{2-d}{2}$, respectively, which are unsurprisingly identical to their values in the linear limit of the pure KPZ equation; see also  Appendix \ref{kpz-lin}. We now study the effects of nonlinear terms on long time and long wavelength scaling behavior of the linear theory by using one loop perturbative RG methods, whose basic steps are outlined above. 

We define dimensionless effective coupling constants $g_1=\frac{\lambda_1^2 D_1}{\nu^3} \tilde{K}_d$, $g_2=\frac{\kappa_1^2 D_T}{\nu_1^2} \tilde{K}_d$ and $g_3=\frac{\kappa_1^2 D_L}{\nu_1^2} \tilde{K}_d$, where $\tilde{K}_d=\frac{\int d\Omega_d}{(2\pi)^d}$ contains the angular contribution coming from the $d$-dimensional volume integral. There are no fluctuation corrections to $D_L$ and $D_T$. Under rescaling of space and time, we rescale the fields in such a way that $D_L$ and $D_T$ do not scale. The ratio $\gamma=\frac{D_T}{D_L}\equiv\frac{g_2}{g_3}$ is marginal in this theory, as we shall see below. Therefore, we do not need to separately study the RG flows of $g_2$ and $g_3$; solving anyone of them suffices, as the other can be eliminated in terms of $\gamma$. For convenience we choose to work with $g_1$ and $g_3$. 
{  Under the rescaling of space, time and the fields { as given} in Appendix~\ref{rescale}, effective coupling $g_1$ scales as $\exp[(2-d)\ell]$, whereas both $g_2$ and $g_3$ scale as $\exp[(2+\mu-d)\ell]$ under these rescalings. Both the pure ($g_1$) and disorder ($g_2$ or $g_3$) nonlinearities 
 na\"ively scale the same way for short-ranged disorder { ($\mu=0$)}, making them compete with each other. Whether the resulting nonequilibrium steady state is controlled by $g_1$, or $g_2$ or $g_3$, or  { all of them}, can only be ascertained by a RG treatment. In contrast, for spatially long-ranged disorders $\mu>0$, near a RG fixed point controlled by $g_2$ or $g_3$, whereas $g_1$ is irrelevant in the RG sense. These can already be seen from the one-loop fluctuation corrections of the model parameters that with short-range quench disorder the corrections originating from the pure nonlinear term $\lambda_1$ have the same  na\"ive infra-red divergence as those originating from the disorder nonlinear term $\kappa_1$. Thus, both types of the fluctuation-corrections must be retained for a RG treatment. In contrast, for long-ranged disorder, the former class is necessarily less infra-red divergent than the latter class (see Appendix \ref{DRG} equations \ref{parameter-correction-kpz}). Therefore, in this case only the latter class of fluctuation-corrections is to be retained, and the former class discarded being less divergent, in the spirit of the RG procedure.}

Note that at 1D the strength $D_T=0$, it implies $g_2=0$ and $\gamma=0$. The relevant one loop Feynman diagrams are shown in Appendix \ref{loop}. The differential RG recursion relations  for the model parameters are:
\begin{subequations}
\begin{align}
& \frac{d D_1}{d\ell}=D_1\left[ z-d-2\chi_h+\frac{g_1}{4}+2g_3 \right],\label{D1-eq}\\
& \frac{d \nu_1}{d\ell}=\nu_1\left[ z-2+ \frac{2-d}{4d}g_1 + 2g_3 \left(\gamma\frac{d-1}{d} + \frac{\mu-d}{2d}\right)\right],\label{nu1-eq}\\
& \frac{d \lambda_1}{d\ell}=\lambda_1\left[ z+\chi_h-2 + 2g_3\left(\gamma\frac{d-1}{d}-\frac{1}{d}\right)\right],\\
& \frac{d \kappa_1}{d\ell}=\kappa_1\left[z-1+\frac{\mu-d}{2}-\frac{2}{d}g_3\right],
\end{align}\label{parameters-kpz flow}
\end{subequations}
along with $dD_L/dl=0=dD_T/dl$.
 The flow equations~(\ref{parameters-kpz flow}) can be used to find the RG flow of the dimensionless couplings $g_1$ and $g_3$. We get
 \begin{widetext}
\begin{subequations}
\begin{align}
 &\frac{d g_1}{d\ell}=g_1\left[ 2-d +\frac{g_1}{4}\left(4- \frac{6}{d}\right) -g_3\left\{2\gamma\left(1-\frac{1}{d}\right) +\left(\frac{3\mu+4}{d}-5\right)\right\}\right],\label{g1-kpz}\\
 &\frac{dg_3}{d\ell}=g_3\left[2+\mu-d-\frac{g_1}{2}\left( \frac{2}{d}-1\right)-g_3\left\{ 4\gamma\left(1-\frac{1}{d}\right)+2 \left(\frac{\mu+2}{d}-1\right) \right\}\right].\label{g3-kpz}
\end{align} \label{gs-eq}
\end{subequations}
\end{widetext}
Below we discuss the scaling behavior for short range and long range correlated disorders separately using the flow equation~(\ref{g1-kpz}) and (\ref{g3-kpz}).

\subsubsection{Short range  disorder}\label{uncorr-kpz}
For the short range disorder case, we set $\mu=0$ as discussed earlier. The RG flow equations of $g_1$ and $g_3$ are
\begin{subequations}
 \begin{align}
  \frac{dg_1}{dl}=& g_1\bigg[ 2-d+\frac{g_1}{4}\left(4- \frac{6}{d}\right) \nonumber\\
    &-g_3\left\{2\gamma\left(1-\frac{1}{d}\right)-\left(5-\frac{4}{d}\right)\right\} \bigg],\label{g1-s-kpz}\\
  \frac{dg_3}{dl}=& g_3[ 2-d+\frac{g_1}{2}\left(1-\frac{2}{d}\right)\nonumber\\
   &-g_3\left\{ 4\gamma\left(1-\frac{1}{d}\right)+2\left(\frac{2}{d}-1\right)\right\} ].\label{g3-s-kpz}
 \end{align} \label{short gs}
 \end{subequations}
 Flow Eqs.~(\ref{g1-s-kpz}) and (\ref{g3-s-kpz}) show that both $g_1$ and $g_3$ have the same critical dimension $2$.  The flow equations in (\ref{short gs}) at 2D have the form
 \begin{align}
  \frac{dg_1}{dl}=g_1\left[\frac{g_1}{4}-g_3(\gamma-3)\right];\, \frac{dg_3}{dl}=-2\gamma g_3^2. \label{g2d-short}
 \end{align}
 In equations (\ref{g2d-short}) (0,0) is the only fixed point. Further, $g_3$ flows to zero in the long RG time limit. In fact, the fixed point (0,0) is {\em stable} along the   $g_3$-direction, whereas {\em unstable} along the 
 $g_1$-direction. We further find that for any ``initial conditions'' $g_1(\ell=0)>0,\,g_3(\ell=0)>0$, the flow ultimately runs away along the $g_1$-axis to infinity suggesting the existence of a perturbatively inaccessible phase. Since $g_3$ flows to zero, we are tempted to speculate that this inaccessible phase is statistically identical to perturbatively inaccessible rough phase of the 2D KPZ equation.
 

 The Gaussian fixed point (0,0) is unstable for dimension at or below $d=2$ and it is the only FP at 2D. For any dimension higher than $d=2$, the Gaussian fixed point is stable and  the nonlinear terms are irrelevant near it. As a result,  the scaling properties are described by linear theory. For $d>2$, there is at least one strong coupling fixed point corresponding to a rough phase, which is perturbatively inaccessible. While this is qualitatively similar to the pure KPZ equation, due to the nonperturbative nature of the rough phase, we cannot tell whether this rough phase is statistically same as the rough phase of $d>2$ pure KPZ equation implying irrelevance of the quenched disorder, or different, in which case, quenched disorder is relevant. In the latter case, we speculate the yet unknown scaling properties are to be parametrised by the parameter $\gamma$. Numerical solutions of (\ref{heq-kpz}), or simulations of suitably constructed discrete models equivalent to (\ref{heq-kpz}) should be able to shed further light on this issue. Unsurprisingly, we further conclude that $d=2$ is the lower critical dimension of (\ref{heq-kpz}), same as for the pure KPZ model.
 
\begin{figure}[!htb]
  \includegraphics[width=0.9\columnwidth]{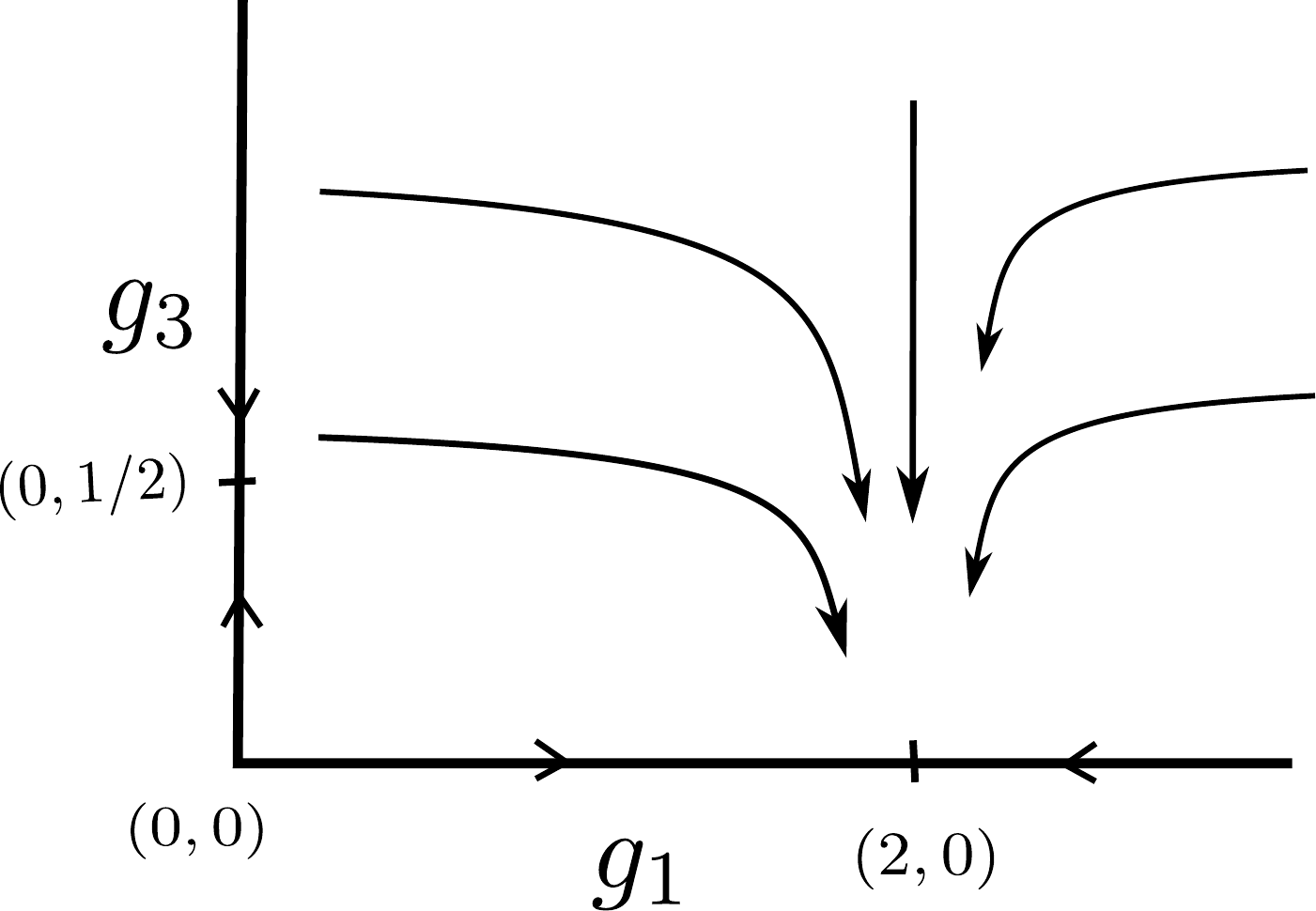}
  \caption{Schematic flow diagram in the $g_1-g_3$ plane for the disordered KPZ equation at 1D, showing the irrelevance of the short ranged quenched disorder (see text).  Here, $(2,0)$ is the stable fixed point.}\label{1d-kpz-phase}
 \end{figure}

 We now briefly consider the scaling properties at $d=1$ by performing the RG at $d=1$, instead of expanding about the critical dimension~\cite{sudip-ckpz,stanley}. Remembering $\gamma=0$, the flow equations  of $g_1$ and $g_3$ are
 \begin{subequations}
 \begin{align}
  \frac{dg_1}{dl}=& g_1\left[ 1-\frac{g_1}{2} +g_3\right],\label{g1-s-kpz-1d}\\
  \frac{dg_3}{dl}=& g_3\left[ 1-\frac{g_1}{2}-2g_3\right].\label{g2-s-kpz-1d}
 \end{align} \label{short gs-1d}
 \end{subequations}
 The Fixed points are found by setting $\frac{dg_1}{dl}=0=\frac{dg_3}{dl}$ and corresponding stability  are given below. 
 \begin{enumerate}
  \item $(0,0)$: globally unstable.
  \item $(0,\frac{1}{2})$:  stable in the $g_3$-direction and  unstable in the $g_1$-direction.
  \item $(2,0)$:  stable in the $g_1$-direction, and marginally stable in the $g_3$-direction. At this stable fixed point, from Eq. (\ref{D1-eq}) and (\ref{nu1-eq}) we get the scaling exponents $z=3/2$ and $\chi_h=1/2$, which characterises the KPZ universality class behavior.
 \end{enumerate}
 The RG flow diagram is pictorially shown in Fig.~\ref{1d-kpz-phase}.
 
 We thus conclude that the short range orientational quenched disorder is irrelevant in (\ref{heq-kpz}), ultimately giving the well-known 1D KPZ universality class.

\subsubsection{Long range correlated disorder}\label{corr-kpz}

In this Section, we study the effects of spatially correlated or long range disorder (i.e., $0<\alpha<d$, i.e., $\mu>0$) on the universal scaling properties of the KPZ equation. 
The flow equations (\ref{g1-kpz}) and (\ref{g3-kpz}) show the critical dimension of $g_1=2$ but for $g_2$ is $2+\mu$. Thus, coupling $g_1$ has a critical dimension lower than $g_3$. Therefore, $g_1$ is subleading to $g_3$, and the leading scaling behavior in the asymptotic long wavelength limit should be controlled by $g_3$ (if the disorder is relevant in the RG sense).  Since $g_1$ is subleading, the renormalised disordered equation motion is invariant under inversion of $h$, a symmetry absent microscopically and also in the rough phase of the pure KPZ equation. Thus, this symmetry under inversion of $h$ appears as an {\em emergent symmetry} of the model in the asymptotic long wavelength limit. The flow Eq.~(\ref{g3-kpz}) for $g_3$ is

\begin{align}
 \frac{dg_3}{dl}=g_3\left[2+\mu-d-g_3\left\{4\gamma(1-\frac{1}{d})+2\left( \frac{\mu+2}{d}-1\right)\right\}\right].\label{g3-kpz-long}
\end{align}
  In order to calculate the scaling exponents at dimensions $d<2+\mu$, we use ${\cal O}(\epsilon_1)$ expansion, where  $d=2+\mu-\epsilon_1$, $\epsilon_1>0$. To the lowest order in ${\cal O}(\epsilon_1)$, the  flow equation (\ref{g3-kpz-long}) for $g_3$ takes the form:
\begin{align}
 \frac{dg_3}{d\ell}=g_3\left[\epsilon_1-\frac{1+\mu}{2+\mu}4\gamma g_3\right]\label{flow-long-kpz}.
\end{align}
 Equation~(\ref{flow-long-kpz}) reveals that  $g_3^*=0$ is an unstable fixed point, but 
 \begin{equation}
 g_3^*=\frac{\epsilon_1}{4\gamma}\frac{2+\mu}{1+\mu} \label{g3-fixed-eps}
 \end{equation}
 for a non-zero $\gamma$ is a stable fixed point. We evaluate the exponents at this stable fixed point. We find the dynamic exponent 
 \begin{equation}
 z=2-\frac{\epsilon_1}{2\gamma}\frac{\gamma(1+\mu)-1}{(1+\mu)},\label{z-long-kpz}
 \end{equation}
 which can be  more or less than $2$, its value in the linear theory, implying that nonlinear disorder effects can induce slower or faster than ordinary diffusive relaxation of fluctuations,  with $\gamma(1+\mu)=1$ is the boundary between the two kinds of behaviour. We further get the roughness exponent 
 \begin{equation}
 \chi_h=\frac{\epsilon_1}{4}[1+\frac{3+\mu}{\gamma(1+\mu)}]-\frac{\mu}{2}. \label{chi-long-kpz}
 \end{equation}
 Thus, both $z$ and $\chi_h$ explicitly vary with $\gamma$. In the limit of $\gamma \rightarrow \infty$, i.e., for $D_T \gg D_L$, $z=2- \epsilon_1/2$ and $\chi_h=\epsilon_1/4 - \mu/2$.  We focus on the $d=2$ case. From the definition of $\epsilon_1$, $d=2$ implies $\epsilon_1=\mu$. From (\ref{chi-ckpz-long}), we then obtain
 \begin{equation}
  \chi_h=\frac{\mu}{4}\bigg[-1+\frac{3+\mu}{\gamma(1+\mu)}\bigg],\label{chi-long-kpz-2d}
 \end{equation}
which can be positive or negative, depending upon $\gamma$. In fact, for a large $\gamma$, $\chi_h<0$. By setting $\chi_h=0$, we get $\gamma_c$, the critical value of $\gamma$:
\begin{equation}
 \gamma_c=(3+\mu)/(1+\mu).
\end{equation}
Now, $\chi_h<0$, which for a fluctuating surface means a ``smooth surface'', means $\langle h^2({\bf x},t)\rangle$ is independent of the system size $L$ for large $L$. This, when $h$ is interpreted as a phase, implies {\em long range orientational order} in 2D, result not possible in equilibrium at 2D due to the well-known Mermin-Wagner theorem (MWT)~\cite{mwt}. Similarly, for an active membrane, this result means the membrane should be statistically flat, an impossibility in equilibrium again due to MWT.  Since in 2D, $\epsilon_1=\mu$ and $\mu=d-\alpha$, the maximum value of $\epsilon_1$ can be 2 ($=\mu$ at 2D), by using this value we find $\chi_h=\frac12[-1+\frac{5}{3\gamma}]$, which can be made negative for $\gamma>5/3$.

 For $\epsilon_1<0$, i.e., $d>2+\mu$, $g_3=0$ is the only fixed point which is stable. We therefore conclude that $d=2+\mu$ is the {\em upper critical dimension} of this model.

 At $d=1$, in which case $D_T=0$ necessarily, giving $\gamma=0$. Then  (\ref{flow-long-kpz}) reads 
\begin{align}
 \frac{dg_3}{d\ell}=g_3\epsilon_1\label{g3-kpz-long-1d}.
\end{align} 
For 1D, we set $\epsilon_1=1+\mu$. The only fixed point here is $g_3^*=0$, which is unstable. The lack of a stable fixed point does not allow us to extract the scaling properties. It is likely to be an artifact of the one loop study.   Notice that at any $d$-dimension for $\gamma\rightarrow 0$, {\em both } $z$ and $\chi_h$ diverge in (\ref{z-long-kpz}) and (\ref{chi-long-kpz}) respectively, which is unphysical. To investigate the $\gamma=0$ case further, we use (\ref{g3-kpz-long}) and perform a fixed dimension RG, which is similar in spirit with the RG for the 1D KPZ equation~\cite{stanley}; see also Refs.~\cite{sudip-ckpz,honko}. In that scheme Eq. (\ref{g3-kpz-long}) gives at $d$-dimension the stable fixed point
\begin{align}
 g_3^*=\frac{d}{2}\cdot\frac{2+\mu-d}{2\gamma(d-1)+2+\mu-d}.\label{g3-fixed-longkpz}
\end{align}
At this fixed point, the scaling exponents can be  calculated from Eqs.~(\ref{D1-eq}) and (\ref{nu1-eq}). We obtain
\begin{eqnarray}
 &&z=2-\frac{(2+\mu-d)[2\gamma(d-1)+\mu-d]}{2[2\gamma(d-1)+2+\mu-d]},\label{z-fixed-longkpz1}\nonumber\\
 &&\chi_h=\frac{2-d}{2}-\frac{(2+\mu-d)[2\gamma(d-1)+\mu+d]}{4[2\gamma(d-1)+2+\mu-d]}.\label{chi-fixed-longkpz}
\end{eqnarray}
 Thus for $\gamma=0$, 
\begin{eqnarray}
 &&z=2-\frac{\mu-d}{2},\\
 &&\chi_h=\frac{2-d}{2} - \frac{\mu+d}{4},
\end{eqnarray}
which are perfectly well-behaved. In particular, at 1D,
we find $g_3^*=1/2, z=2-\frac{\mu-1}{2}$ and $\chi_h=1-\frac{\mu+1}{4}$.  Further, using (\ref{chi-fixed-longkpz}) we find in 2D $\chi_h=-\frac{\mu(2\gamma+2+\mu)}{4(2\gamma+\mu)}$ which gives a smooth surface whatever value of $\mu$ and $\gamma$. Generally then,
regardless of the scheme we used, we uncover explicit $\gamma$-dependence of $z$ and $\chi_h$. This clearly shows parametrisation of the universality classes by $\gamma$.


 In general, an alert reader will easily notice that the roughness exponent in Eq.~(\ref{chi-long-kpz}) from $\epsilon_1$ expansion RG, or in Eq.~(\ref{chi-fixed-longkpz}) from fixed dimension RG method reveals that by tuning $\mu$, it can be made positive or negative for $d<2+\mu$. Thus, by changing $\mu$, the surface can be made rough ($\chi_h>0$) or smooth ($\chi_h<0$). The transition line in the $\mu-d$ plane can be found by setting $\chi_h=0$. By using (\ref{chi-long-kpz}), we find the equation for the line
\begin{align}
 d=2+\mu-\frac{2\mu\gamma(1+\mu)}{\gamma(1+\mu)+3+\mu},\label{pt-ep-kpzlong}
\end{align}
in the $\epsilon_1$-expansion calculation,
and by using (\ref{chi-fixed-longkpz}), the line is given by
\begin{align}
 2-d=\frac{(2+\mu-d)[2\gamma(d-1)+\mu+d]}{2[2\gamma(d-1)+2+\mu-d]},\label{pt-fd-kpzlong}
\end{align}
in the fixed dimension RG calculation.  On the line $\chi_h=0$, the surface is {\em logarithmically rough}. The dynamic exponent $z$ on that line unsurprisingly depends on the precise equation of the line $\chi_h=0$. We show for two differenty method:  $z=2+\frac{\mu[1-\gamma(1+\mu)]}{3+\mu+\gamma(1+\mu)}$ by using eq. (\ref{pt-ep-kpzlong}) in $\epsilon_1$ expansion case and $z=d+\frac{d(2+\mu-d)}{2\gamma(d-1)+2+\mu-d}$ by using (\ref{pt-fd-kpzlong}) in fixed dimension RG analysis.

We plot the values of the scaling  exponents  as obtained from fixed dimension RG and $\epsilon_1$ expansion as functions of $\gamma$ in Fig.~\ref{exponent-longkpz} below.
\begin{figure}[!tb]
 \includegraphics[width=\columnwidth]{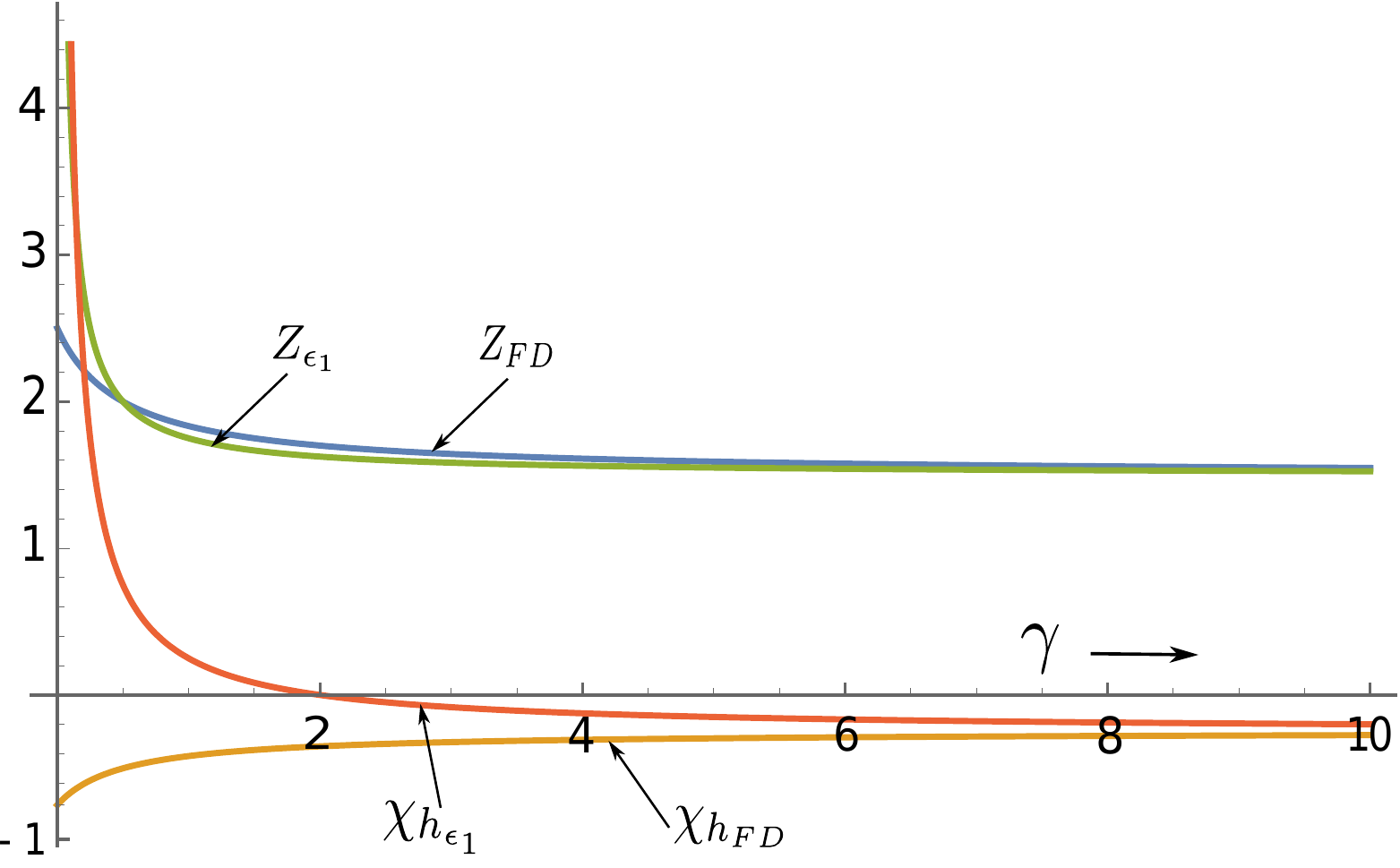}
 \caption{(color online) A plot of the scaling exponents of the quenched disordered KPZ equation, comparing between the results found from one-loop $\epsilon_1$ expansion and fixed dimension RG, showing the $\gamma$-dependence of the scaling exponents. Here, suffix $\epsilon_1$ indicates scaling exponents obtained from an $\epsilon_2$-expansion, whereas suffix $FD$ indicates fixed dimension RG results. These are plotted for $d=2$ and $\mu=1$. }\label{exponent-longkpz}
\end{figure}

It is theoretically interesting to calculate the scaling exponents at the critical dimension $d=2+\mu$. The flow equation of $g_3$ is at $d=2+\mu$ reads [see Eq.~(\ref{flow-long-kpz})]
\begin{align}
 \frac{dg_3}{d\ell}=-\frac{1+\mu}{2+\mu}4\gamma g_3^2.\label{flow-long-dc-kpz}
\end{align}
Equation~(\ref{flow-long-dc-kpz}) shows that $g_3(\ell)$ is marginally irrelevant; $g_3(\ell)$  flows towards $g_3^*=0$ with increasing of $\ell$. Coupling $g_3(\ell)$ takes the form $g_3(\ell)=\frac{g_3(0)}{1+g_3(0)\ell4\gamma (\frac{1+\mu}{2+\mu})}$, and for large renormalisation group time i.e., $\ell\rightarrow\infty$, $g_3(\ell)\simeq\frac{2+\mu}{4\gamma(1+\mu)}\frac{1}{\ell}$. This shows that time-scale $t$ no longer shows simple scale with length-scale $r$, giving breakdown of conventional dynamic scaling
\begin{equation}
t\sim r^2(\log(r/a_0))^{-\frac{\gamma(1+\mu)-1}{2\gamma(2+\mu)}}. 
\end{equation}
Thus, the extent of breakdown of dynamic scaling depends on $\gamma$,  and can be faster or slower than ordinary diffusion. 
{ We now calculate the variance of $h$ at the critical dimension $d=2+\mu$. We get
\begin{align}
 \langle h^2({\bf x},t)\rangle\simeq\int_{1/L}^{1/a_0} {d^{2+\mu}{\bf q}}~ q^{-2}[\log(1/q)]^{-\frac{\gamma(1+\mu)-(3+\mu)}{2\gamma(2+\mu)}}. \label{h2-kpzlong}
\end{align}
{ Here, $L$ and $a_0$ are linear system size and small-scale cutoff, respectively.}
Clearly, $\langle h^2({\bf x},t)\rangle$ is bounded: it does not diverge for large $L$.
}
For $d>2+\mu$, i.e., above the upper critical dimension, since $g_3=0$ is the stable FP, the asymptotic long wavelength limit scaling is identical to the linear theory.
 
 We thus find that with short range disorder although the disorder coupling constant $g_3$ is na\"ively as relevant as the coupling $g_1$ of the pure KPZ equation, in a one-loop theory $g_3$ is irrelevant in the RG sense at all dimensions. For long range correlated disorder $g_3$ is relevant while $g_1$ becomes irrelevant, resulting into scaling exponents depending explicitly on $\gamma$.  { Lastly, for $\mu <0$ quenched disorder is irrelevant in all dimensions. Hence, the model belongs to the standard KPZ universality class. In fact, this model with $\mu<0$ is statistically identical to $\mu=0$ in the long wavelength limit.}
 
\begin{figure}[!htb]
  \includegraphics[width=0.9\columnwidth]{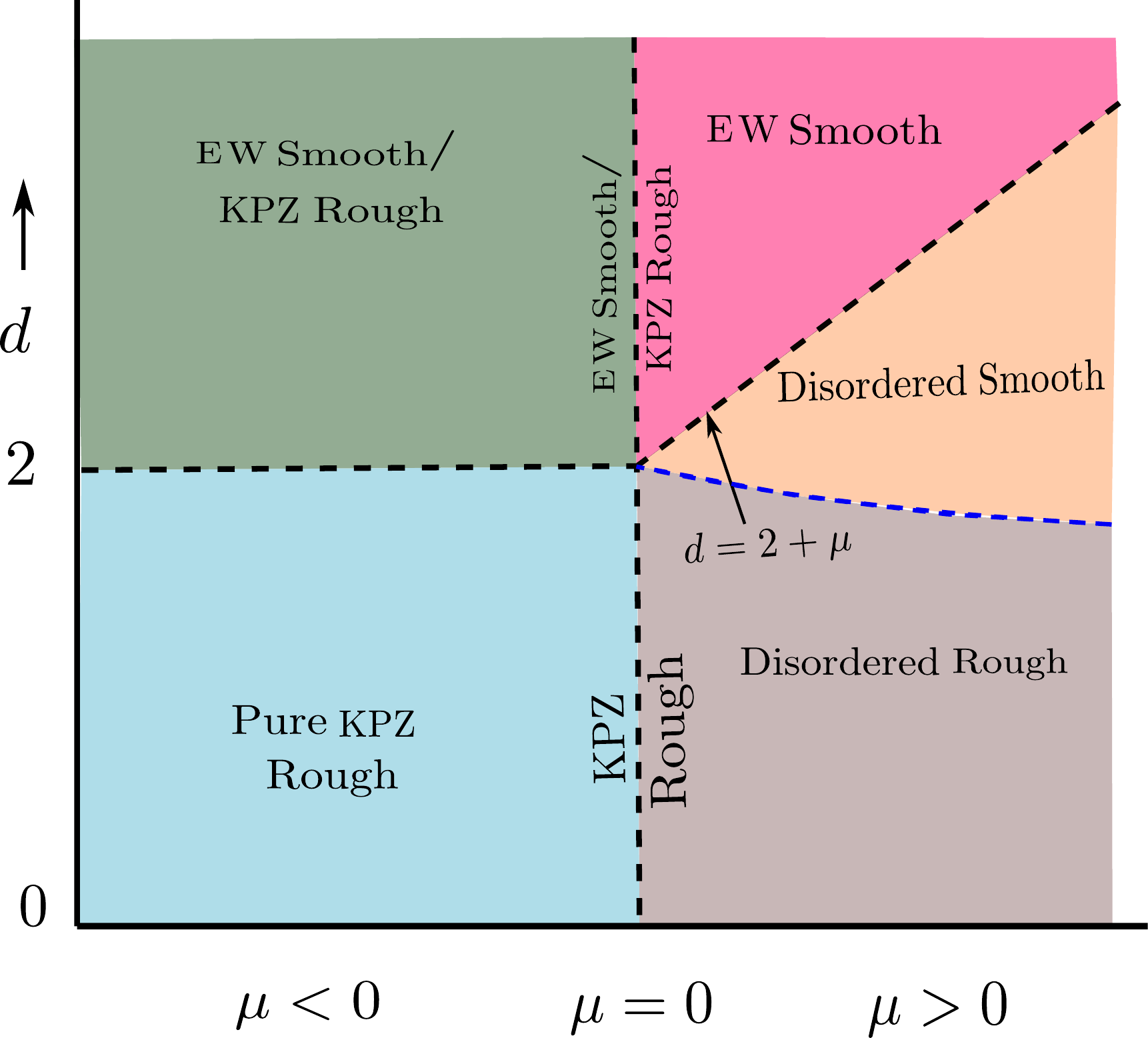}
  \caption{(color online) Schematic phase diagram of the quenched disordered KPZ equation in $\mu-d$ plane. The broken blue line corresponds to $\chi_h=0$, dividing the disorder-induced rough and smooth phases, and depends on $\gamma$; here, this line is schematically drawn for $\gamma=1$).}\label{kpzphase}
 \end{figure}

 We present the phase diagram of the disordered KPZ equation in the $\mu-d$ plane showing the possible phases in Fig.~\ref{kpzphase}

\subsection{Quenched disordered CKPZ equation}\label{ckpz-rg}

We now investigate the universal scaling properties of the quenched disordered CKPZ equation. As in our study of the disordered KPZ equation above, we consider both short-ranged and long-ranged disorders separately. In the linear limit of (\ref{heq-ckpz}), we find
 dynamic exponent $z=4$ and roughness exponent $\chi_h=\frac{2-d}{2}$, which are unsurprisingly identical to the scaling of the linearlised  pure CKPZ equation; see Appendix \ref{ckpz-lin} for detailed calculations. The nonlinear terms can affect the linear theory theory scaling, if they are relevant in a RG sense. For instance, the pure CKPZ equation has an upper critical dimension two, meaning for dimension $d<2$, the nonlinear effects modify the linear theory scaling in the asymptotic long wavelength limit. In the disordered case, as for the quenched disordered KPZ equation, the pertinent questions are, whether disorder is relevant in the RG sense, and if so, what are scaling properties of the resulting universality class. We systematically address these issues by using one-loop perturbative RG, whose details are available in the Appendix~\ref{DRG}.
 

 Similar to our studies above on the quenched disordered KPZ equation, we define two dimensionless coupling constants $\varTheta_1=\frac{\lambda_2^2 D_2}{\nu_2^3} \tilde{K}_d$, $\varTheta_2=\frac{\kappa_2^2 D_T}{\nu_2^2} \tilde{K}_d$ and $\varTheta_3=\frac{\kappa_2^2 D_L}{\nu_2^2} \tilde{K}_d$. Further, we set $D_L$ and $D_T$ do not scale, accordingly we define a ratio $\gamma=\frac{D_T}{D_L}\equiv\frac{\varTheta_2}{\varTheta_3}$. Note that at 1D the strength $D_T=0$, which implies $\gamma=0$, as before. Furthermore, na\"ive rescaling of space, time and the fields, as shown in Appendix~\ref{rescale} show that the under these rescalings, effective couplings $\varTheta_1$ scales as $\exp[(2-d)\ell]$, whereas both $\varTheta_2$ and $\varTheta_3$ scale as $\exp[(2+\mu-d)\ell]$.  { As in the RG treatment of the disordered KPZ equation, in the RG analysis for the CKPZ equation,}   both the pure ($\varTheta_1$) and disorder ($\varTheta_2$ or $\varTheta_3$) nonlinearities 
 na\"ively scale the same way for short-ranged disorder, making them compete with each other and control the nonequilibrium steady states.  In contrast, for spatially long-ranged disorders $\mu>0$, $\varTheta_2$ or $\varTheta_3$ control the behavior of stedy states, $\varTheta_1$ is irrelevant in the RG sense .   { The one-loop fluctuation-corrections generating from pure $\lambda_2$ and disorder $\kappa_2$ nonlinearities show these two types of infrared divergences (see Appendix \ref{DRG} equations \ref{parameter-correction-ckpz}). With  short-range disorder, both types have { the} same divergences, but for long range disorder the leading divergences  { originate} from { the} disorder { vertices}}. Needless to say, as for the disordered KPZ equation studied above, in all these cases the parameter $\gamma$ is marginal, and has no fluctuation-corrections. This means there is no need to study the flows of $\varTheta_2$ and $\varTheta_3$ separately; it is enough to study the flow of one of them parametrised by $\gamma$. In what follows below, we choose to work with $\varTheta_1$ and $\varTheta_3$.

The RG recursion relation of the model parameters for general $\mu\geq 0$ are (see Appendix~\ref{DRG} for more details):
\begin{subequations}
\begin{align}
& \frac{d D_2}{d\ell}=D_2\left[ z-d-2-2\chi_h  \right],\label{D2flow}\\
& \frac{d \nu_2}{d\ell}=\nu_2\left[ z-4+ \frac{4-d}{4d}\varTheta_1 + 2\varTheta_3 \left(\gamma\frac{d-1}{d} + \frac{\mu-d}{2d}\right)\right],\label{nu2flow}\\
& \frac{d \lambda_2}{d\ell}=\lambda_2\left[ z+\chi_h-4 + 2\varTheta_3\left(\gamma\frac{d-1}{d}-\frac{1}{d}\right)\right],\label{lambda2flow}\\
& \frac{d \kappa_2}{d\ell}=\kappa_2\left[z-3+\frac{\mu-d}{2}-\frac{2}{d}\varTheta_3\right].\label{kappa2flow}
\end{align}\label{parameters flow ckpz}
\end{subequations}

The flow equations of the dimensionless coupling constants $\varTheta_1$ and $\varTheta_3$ are then calculated by using (\ref{parameters flow ckpz}). These are given by
\begin{widetext}
\begin{subequations}
\begin{align}
 &\frac{d \varTheta_1}{d\ell}=\varTheta_1\left[ 2-d - \frac{4-d}{4d}3\varTheta_1-\varTheta_3\left(2\gamma\frac{d-1}{d} +\frac{4+3\mu-3d}{d}\right)\right],\label{theta1-eq}\\
 &\frac{d\varTheta_3}{d\ell}=\varTheta_3\left[2+\mu-d-\frac{4-d}{2d}\varTheta_1-\varTheta_3\left\{\frac{d-1}{d}4\gamma+2\left(\frac{2+\mu-d}{d}\right)\right\}\right].\label{theta3-eq}
\end{align} \label{theta-eq}
\end{subequations}
\end{widetext}

We separately discuss the short and long range disorder cases below.

\subsubsection{Short range disorder}\label{uncorr-ckpz}

When the disorder is short-ranged, we set $\mu=0$. 
The flow equations of couplings $\varTheta_1$ and $\varTheta_3$
 for the short-ranged disorder case can be obtained from (\ref{parameters flow ckpz}). These are
\begin{subequations}
\begin{align}
 &\frac{d \varTheta_1}{d\ell}=\varTheta_1\left[ 2-d - \frac{4-d}{4d}3\varTheta_1-\varTheta_3\left(2\gamma\frac{d-1}{d} +\frac{4-3d}{d}\right)\right],\label{theta1-s-ckpz}\\
 &\frac{d\varTheta_3}{d\ell}=\varTheta_3\left[2-d-\frac{4-d}{4d}2\varTheta_1-2\varTheta_3\left\{2\gamma\frac{d-1}{d}+\frac{2-d}{d}\right\}\right].\label{theta3-s-ckpz}
\end{align}\label{theta-s}
\end{subequations}
Flow equations (\ref{theta1-s-ckpz}) and (\ref{theta3-s-ckpz}) show that $\varTheta_1$ and  $\varTheta_3$ both have the same critical dimension $2$, as already argued by using na\"ive rescaling of space, time and the fields. Further, the Gaussian fixed point (0,0) is globally unstable below 2D, whereas this is the only stable fixed point  at or above 2D.   That implies that $d=2$ is the upper critical dimension for this model, same as pure CKPZ; at 2D, couplings $\varTheta_1$ and $\varTheta_3$ are {\em marginally irrelevant}. Therefore, the linear theory scaling is expected to be modified by the nonlinear effects at or below 2D, whereas for $d>2$ the nonlinear couplings are irrelevant and the linear theory holds in the asymptotic long wavelength limit.  To the lowest order in $\epsilon_2\equiv 2-d$, the flow equations (\ref{theta1-s-ckpz}) and (\ref{theta3-s-ckpz}) reduce to

\begin{subequations}
\begin{align}
 &\frac{d \varTheta_1}{d\ell}=\varTheta_1\left[ \epsilon_2 - \frac{3\varTheta_1}{4}-\varTheta_3\left(\gamma -1\right)\right].\label{theta1-1d-s}\\
 &\frac{d\varTheta_3}{d\ell}=\varTheta_3\left[\epsilon_2-\frac{\varTheta_1}{2}-2\gamma\varTheta_3\right].\label{theta3-1d-s}
\end{align}\label{theta-1d}
\end{subequations}

Equations~(\ref{theta1-1d-s}) and (\ref{theta3-1d-s}) can be used to calculate the { fixed points} $(\varTheta_1^*, \varTheta_2^*)$ by setting $\frac{d\varTheta_1}{d\ell}=0=\frac{d\varTheta_3}{d\ell}$. We find 


\begin{enumerate}
 \item  Fixed point 1: (0,0) is globally  unstable.  
 \item Fixed point 2: (0,$\epsilon_2/2$)  unstable in the $\varTheta_1$-direction, but stable in the $\varTheta_3$-direction.
 \item Fixed point 3: ($4\epsilon_2/3,0$) unstable in the $\varTheta_3$-direction, but stable in the $\varTheta_1$-direction.
 \item Fixed point 4: The nontrivial fixed point ($\frac{2\epsilon_2(1+\gamma)}{2\gamma+1}, \frac{\epsilon_2}{2(2\gamma+1)}$): stable along both the $\varTheta_1$ and $\varTheta_2$ directions, i.e., globally stable.\\
 \end{enumerate}

\begin{figure}[!htb]
 \includegraphics[width=0.9\columnwidth]{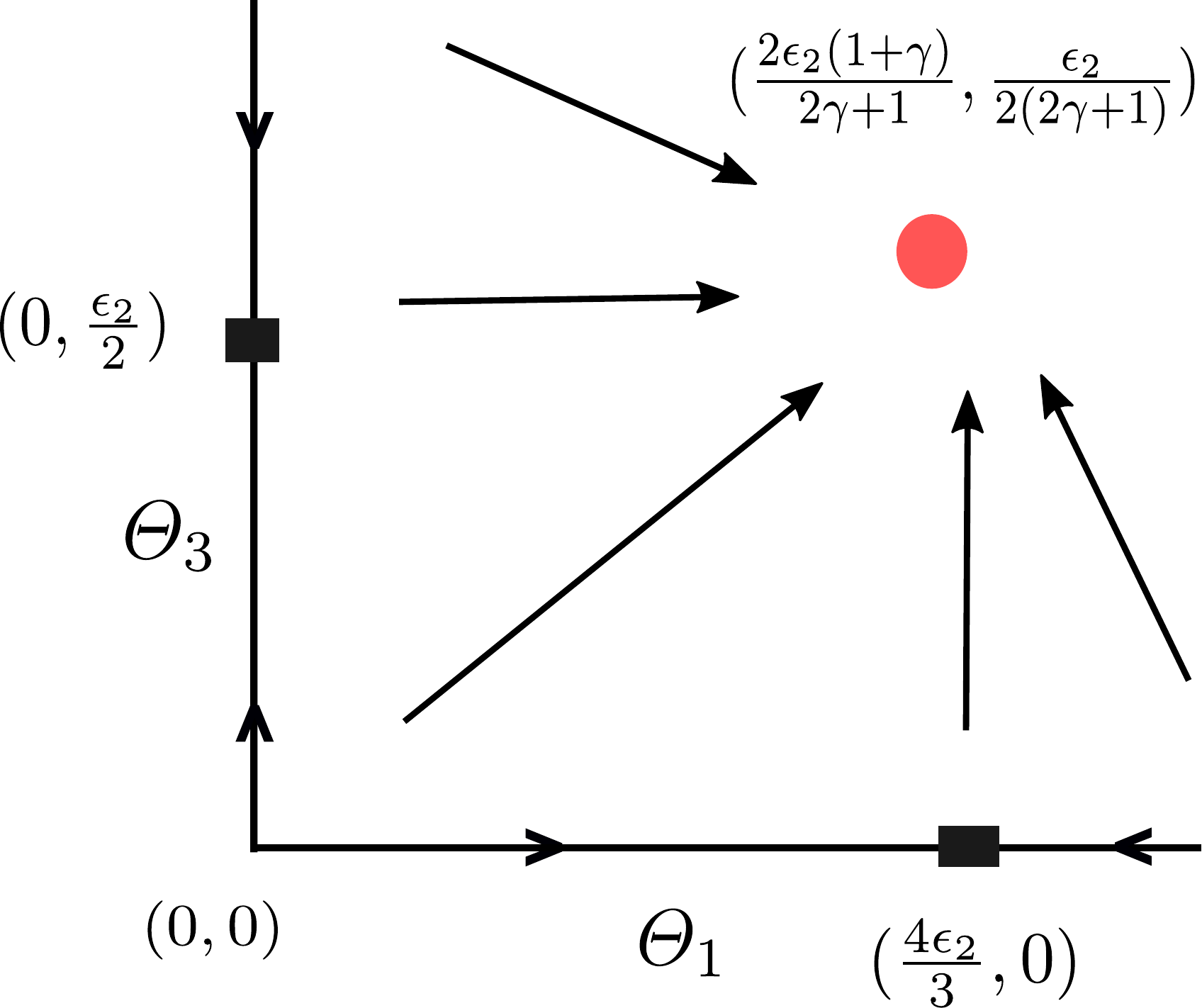}
\caption{(color online) Schematic flow diagram at dimension $d=2-\epsilon_2,\, \epsilon_2>0$ (except 1D) for the short range quenched disordered CKPZ equation. The small, red circle is the nontrivial fixed point, which is globally stable. }\label{short-ckpz-less2d}
\end{figure}

 The fixed points and the RG flow lines are shown schematically in Fig.~\ref{short-ckpz-less2d}.

 The globally stable fixed point describes the stable nonequilibrium steady state of the model. It can be used to  calculate the associated scaling exponents. We find that the dynamic exponent 
 \begin{equation}
 z=4-\frac{\epsilon_2\gamma}{2\gamma+1},
 \end{equation}
  and the roughness exponent 
  \begin{equation}
  \chi_h=\frac{\epsilon_2(1+\gamma)}{2(2\gamma+1)}.
  \end{equation}
  { From this above expression $\chi_h>0$ since $\epsilon_2>0$. { Thus}, it is  a 
  short-range disorder induced rough phase (``SR rough'')}. We note that as $\gamma\rightarrow 0$, both $z$ and $\chi_h$ approach their linear theory values, i.e., $z=4,\,\chi=\epsilon_2/2$, an unexpected result.  
  In the same way, we note that at 1D, $\epsilon_2=1$ and $\gamma=0$, giving  $z=4$, $\chi_h=\frac{1}{2}$, identical to the linear theory results! In the absence of any general symmetry argument to render the fluctuation corrections to vanish or turn irrelevant, we believe this is fortuitous, an artifact of the one-loop perturbation theory. Higher order perturbation theory or numerical studies should be useful to obtain better quantitative estimates of the exponents in the limit $\gamma\rightarrow 0$.  We re-analyse the 1D case by using a fixed dimension RG scheme, which is similar in spirit with what one does for the 1D KPZ equation~\cite{stanley}. To do this, we use  the flow equations (\ref{theta1-s-ckpz}) and (\ref{theta3-s-ckpz}) directly, and set $d=1$.
  
  
   The resulting flow  equations 
 \begin{align}
  &\frac{d\varTheta_1}{d\ell}=\left[1-9\varTheta_1/4-\varTheta_3\right],\\
  &\frac{d\varTheta_3}{d\ell}=\left[1-3\varTheta_1/2-2\varTheta_3\right].
 \end{align}
Solving these two equations, $(1/3,1/4)$ is only stable FP and finds the exponents $z=4, \chi_h=1/2$, again same as the linear theory results, offering no further insight into the unexpected appearance of the linear theory results.

The fixed point analysis of the flow equations in 2D, which is the physically relevant dimension as its pure counterpart is, needs to be done separately, as 2D is the critical dimension. At $d=2$ i.e., $\epsilon_2=0$.
The resulting flow equations of $\varTheta_1$ and $\varTheta_3$ are
\begin{subequations}
\begin{align}
 &\frac{d \varTheta_1}{d\ell}= -\varTheta_1\left[ \frac{3\varTheta_1}{4}+\varTheta_3\left(\gamma-1\right)\right].\label{theta1-2d-ckpz}\\
 &\frac{d\varTheta_3}{d\ell}= -\varTheta_3\left[\frac{\varTheta_1}{2}+2\gamma\varTheta_3\right].\label{theta3-2d-ckpz}
\end{align}
\end{subequations}
 Thus (0,0) is the {\em only } fixed point. Whether it is a stable or unstable fixed point can be found only by solving the flow Eqs.~(\ref{theta1-2d-ckpz}) and (\ref{theta3-2d-ckpz}).
In order to do so, we first define a ratio $\beta=\frac{\varTheta_3}{\varTheta_1}$. The flow equation of $\beta$ is 
\begin{align}
&\frac{d\beta}{d\ell}=\beta \varTheta_1 \left[\frac{1}{4}-\beta(1+\gamma)\right].\label{beta}
\end{align}
 Equation~(\ref{beta}) shows that
 \begin{equation}
 \beta^*=\frac{1}{4(1+\gamma)}=\frac{\varTheta_3(\ell)}{\varTheta_1(\ell)} \label{beta-fp}
 \end{equation}
 gives the fixed point of (\ref{beta}), which in turn
 gives the {\em separatrix}, such that all initial conditions $\varTheta_1(\ell=0),\,\varTheta_3(\ell=0)$ maintaining $\varTheta_3(\ell=0)/\varTheta_1(\ell=0)=\beta^*$ will continue to maintain it under the RG transformations. Furthermore, the separatrix can be shown to be {\em stable} or {\em attractive}: We write $\beta=\beta^*+\delta\beta$. To the linear order in $\delta\beta$, we find
 \begin{equation}
  \frac{d\delta\beta}{\delta\ell}=-\beta^*\varTheta_1(1+\gamma)\delta\beta,
 \end{equation}
showing the attractive nature of the separatrix.  Interestingly, as $\gamma\rightarrow \infty$, $\beta^*\rightarrow 0$, meaning that the separatrix coincides with the $\varTheta_1$-axis in the $\varTheta_1-\varTheta_3$ plane in that limit.
 Using the value of $\beta^*$, we can further show that {
 \begin{subequations}
 \begin{align}
 &&\frac{d\varTheta_1(\ell)}{d\ell} =-\frac{\varTheta_1^2(1+2\gamma)}{2(1+\gamma)}<0,\\
 &&\frac{d\varTheta_3(\ell)}{d\ell} =-\varTheta_3^22(1+2\gamma)<0,
 \end{align}
 \end{subequations}
 }
 giving (0,0) as the {\em globally stable} fixed point. This { then} means $\varTheta_1$ and $\varTheta_3$ are {\em marginally irrelevant}, with both of them eventually flowing to zero in the asymptotic long wavelength limit. Nonetheless, they do so slowly enough to allow for infinite renormalisation of $\nu_2$, leading to logarithmic modulations of the linear theory scaling. We work this out explicitly below. 
 
\begin{figure}[!htb]
 \includegraphics[width=0.8\columnwidth]{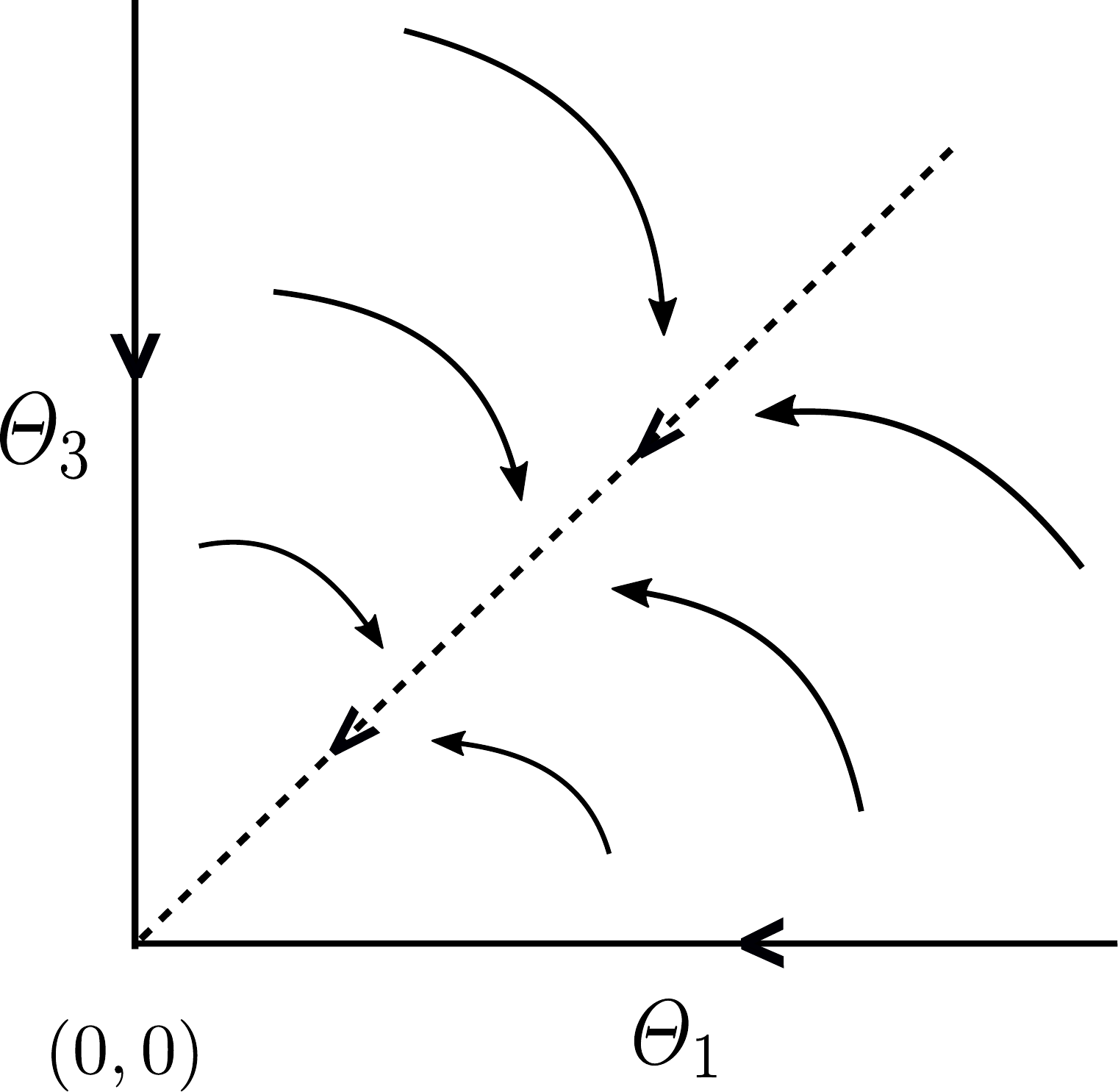}
\caption{Schematic RG flow diagram in 2D ($\epsilon_2 = 0$) for the short-range
quenched disordered CKPZ equation. The origin $(0,0)$ is the only fixed point, which is stable. The
inclined straight line is the separatrix having a $\gamma$-dependent slope given by Eq.~(\ref{beta-fp}).}\label{short-ckpz-2d}
 \end{figure}

 Since the separatrix is attractive, we can use $\varTheta_3(l)=\beta^*\varTheta_1(l)$ in the long wavelength limit in the flow equation (\ref{theta1-2d-ckpz}). We can then solve the resulting effective flow equation for $\varTheta_1(l)$
 giving
 \begin{equation}
  \varTheta_{1l}=\frac{2(1+\gamma)}{1+2\gamma}\frac{1}{\ell},\;\varTheta_{3l}=\frac{1}{2(1+2\gamma)}\frac{1}{\ell},\label{stable2d-l}
 \end{equation}
is the asymptotic limit $\ell \rightarrow \infty$. Since $\ell$ is the logarithm of a physical length scale, we note that both $\varTheta_1(\ell)$ and $\varTheta_3(\ell)$ approach zero logarithmically slowly. Thus, an arbitrary initial condition $\varTheta_1(\ell=0),\,\varTheta_3(\ell=0)$ not only {\em flows to} the origin, it also {\em flows towards} the attractive separatrix in course of its flow to the origin under successive RG transformations. The RG flows in 2D are schematically shown in Fig.~\ref{short-ckpz-2d}.

 %
 We can now find the scale-dependent $\nu_2(\ell)$ by assuming $z=4,\,\chi_h={ 0}$. This gives
 
 \begin{align}
  \nu_2(\ell)\simeq\nu_2(0)\ell^{\frac{\gamma}{2\gamma+1}}.
 \end{align}
However, $D_2(\ell)\simeq D_2(0)$, since it has no relevant one-loop fluctuation correction.

 These can be used to obtain the variance
 \begin{align}
  \langle h^2({\bf x},t) \rangle \sim \left[\log (L/a_0) \right]^{1-\frac{\gamma}{1+2\gamma}},\label{log-rough}
 \end{align}
 where $L$ and $a_0$ are linear system size and small-scale cutoff, respectively. { The logarithmic dependence of $ \langle h^2({\bf x},t) \rangle$ on the system size $L$ in (\ref{log-rough}) implies that at $d=2$, the disordered CKPZ surface is {\em logarithmically rough} (``log rough''). }  We further see that in the renormalised theory, time-scale $t$ and length-scale $r$ are no longer simply related by a dynamic scaling exponent. We in fact find
 \begin{equation}
  t\sim r^4 [\log(r/a_0)]^{-\frac{\gamma}{1+2\gamma}}
 \end{equation}
for large $r$, signifying breakdown of conventional dynamic scaling due to logarithmic modulations. The dynamics is logarithmically slower than that in the linear theory by an extent that depends upon $\gamma$. In the limit $\gamma=0$ ($D_T=0$), $t\sim r^4$, same as in the linear theory, which is likely an artifact of our low order perturbation theory; as $\gamma$ rises, the dynamics gets slower initially, ultimately saturating at $t\sim r^4 \frac{1}{\sqrt{\log (r/a_0)}}$ as $\gamma\rightarrow \infty$, i.e., $D_T\gg D_L$. { In the same limit by using (\ref{log-rough}), we obtain $\langle h^2({\bf x},t) \rangle \sim \sqrt{\log (L/a_0) }$.}


 \subsubsection{Long-range correlated disorder}\label{corr-ckpz}

We now study the scaling properties in the presence of long range quenched disorder, i.e., $\alpha<d$, or $\mu>0$. 
The flow equations (\ref{theta1-eq}) and (\ref{theta3-eq}) show that the critical dimensions of $\varTheta_1=2$ but for $\varTheta_3$ is $2+\mu>0$. Here $\varTheta_1$ has critical dimension lower than $\varTheta_3$. Therefore, near a fixed point controlled by $\varTheta_3$, $\varTheta_1$ is subleading to $\varTheta_3$, and the leading scaling behavior is described by $\varTheta_3$. This is similar to our analysis for the long range quenched disordered KPZ equation above.  Again, as in the disordered KPZ equation, the disordered CKPZ equation with long range disorder has invariance under inversion of $h$ as an emergent symmetry in the long wavelength limit. Hence, Eq.~(\ref{theta3-eq}) reduces to
\begin{align}
 \frac{d\varTheta_3}{dl}=\varTheta_3\left[2+\mu-d-\varTheta_3\left\{4\gamma(1-\frac{1}{d})+2\left( \frac{\mu+2}{d}-1\right)\right\}\right].\label{theta3-ckpz-long}
\end{align}
Notice that Eq.~(\ref{theta3-ckpz-long}) is identical to the flow equation (\ref{g3-kpz-long}) in Sec.~\ref{corr-kpz}, and hence, the ensuing analysis runs exactly parallel to the one below Eq.~(\ref{g3-kpz-long}) in Sec. \ref{corr-kpz}.
To proceed further, we set   $d=2+\mu-\epsilon_2$ where $\epsilon_2>0$ and extract the fixed point and the associated scaling exponents to ${\cal O}(\epsilon_2)$. The flow equation of $\varTheta_3$ reads 
\begin{align}
 \frac{d\varTheta_3}{d\ell}=\varTheta_3\left[\epsilon_2-\frac{1+\mu}{2+\mu}4\gamma\varTheta_3\right]\label{flow-long-ckpz}.
\end{align}
Unsurprisingly, Eq.~(\ref{flow-long-ckpz}) is same as Eq.~(\ref{flow-long-kpz}) in Sec.~\ref{corr-kpz}.
Equation~(\ref{flow-long-ckpz}) gives  $\varTheta_3=0$ is an unstable fixed point, but $\varTheta_3=\varTheta_3^*\equiv\frac{\epsilon_2}{4\gamma}\frac{2+\mu}{1+\mu}$ is a stable {\em fixed point} for any $\gamma>0$. 
As before, the scaling exponents are calculated at the stable  fixed point $\varTheta_3^*$. We find the dynamic exponent
\begin{equation}
z=4-\frac{\epsilon_2}{2\gamma}\frac{\gamma(1+\mu)-1}{1+\mu}
\end{equation}
which is less than $4$. Thus, the dynamics is faster than in the linear theory. Further, the roughness exponent 
\begin{equation}
\chi_h=\frac{\epsilon_2}{4}[1+\frac{1}{\gamma(1+\mu)}]-\frac{\mu}{2}. \label{chi-ckpz-long}
\end{equation}
Clearly, both $z$ and $\chi$ depends explicitly on and vary continuously with $\gamma$.  As $\gamma\rightarrow \infty$, $z=4-\epsilon_2/2,\;\chi_h=\epsilon_2/4-\mu/2$. On the other hand, if $\gamma\rightarrow 0$, i.e., $D_T\ll D_L$, both $z$ and $\chi_h$ diverge, an unexpected result reminiscent of what we have found in the analogous analysis for the long range disordered KPZ equation.  At 1D, necessarily $\gamma=0$, the flow of $\varTheta_3$ reads same as (\ref{g3-kpz-long-1d})
\begin{align}
 \frac{d\varTheta_3}{d\ell}=\varTheta_3\epsilon_2.\label{theta3-long-1d}
\end{align}
Equation~(\ref{theta3-long-1d}) surprisingly has no stable FP, so it cannot describe stable scaling behaviour at 1D, which we believe just as an artifact of our one-loop calculations. 
To investigate this further, we perform a fixed dimension RG akin to the 1D KPZ equation~\cite{stanley}.
 In this scheme  flow Eq.~(\ref{theta3-ckpz-long}) has a fixed point $\varTheta_3^*=\frac{d}{2}\cdot\frac{2+\mu-d}{2\gamma(d-1)+2+\mu-d}$ at a (fixed) dimension $d$, which is a stable fixed point. We calculate  the exponents at this fixed point are
 \begin{subequations}
\begin{align}
 &z=4-\frac{(2+\mu-d)[2\gamma(d-1)+\mu-d]}{2[2\gamma(d-1)+2+\mu-d]},\label{z-fixed-ckpzlong}\\
 &\chi_h=\frac{2-d}{2}-\frac{(2+\mu-d)[2\gamma(d-1)+\mu-d]}{4[2\gamma(d-1)+2+\mu-d]}.\label{chi-fixed-ckpzlong}
\end{align}
\end{subequations}
Thus the scaling exponents depend explicitly on $\gamma$.
These results reveal the generally expected existence of a stable scaling regime parametrised by $\gamma$ different from the linear theory scaling.

\begin{figure}[!htb]
 \includegraphics[width=\columnwidth]{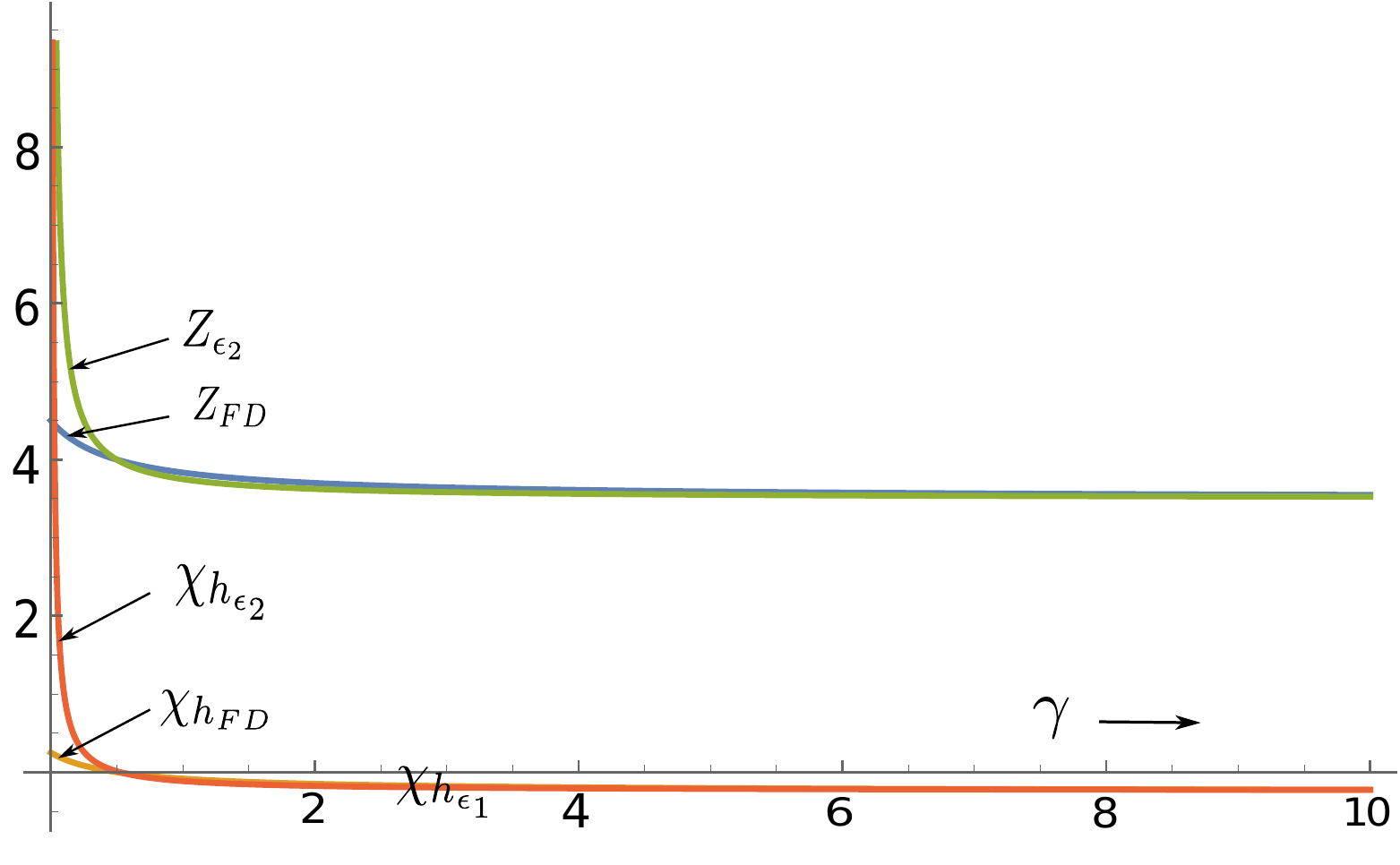}
 \caption{(color online) The scaling exponents are plotted as functions of $\gamma$ to compare between the results found from one-loop $\epsilon_2$ expansion RG and fixed dimension RG. Here, suffix $\epsilon_2$ indicates scaling exponents obtained from an $\epsilon_2$-expansion, whereas suffix $FD$ indicates fixed dimension RG results. We have set $d=2,\, \mu=1$.}\label{exponent-draw-long-ckpz}
\end{figure}

We give a plot of the scaling exponents, showing their dependence on the disorder parameter $\gamma$ in Fig.~\ref{exponent-draw-long-ckpz}.

 For $d=1$ with $\gamma=0$, the value of the fixed point is $\varTheta_3=1/2$ and exponents are $z=4-\frac{\mu-1}{2}$ and $\chi_h=\frac{1}{2}-\frac{\mu-1}{4}$ using the values of (\ref{z-fixed-ckpzlong}) and (\ref{chi-fixed-ckpzlong}). 
 
 Similar to the disordered KPZ equation with long range quenched disorder, both smooth and rough phases are possible for $d<2+\mu$, as can be seen from Eq.~(\ref{chi-ckpz-long}), obtained from an $\epsilon_1$ expansion RG, and from Eq.~(\ref{chi-fixed-ckpzlong}), obtained by using a fixed dimension RG method. As $\gamma$ is varied, the system can undergo a transition between a smooth with $\chi_h<0$ and a rough $\chi_h>0$ phase. The transition line in the $\mu-d$ plane
 can be found from the condition $\chi_h=0$, and is parametrised by $\gamma$. The equation of this line obtained from an $\epsilon_2$ expansion RG  is given by
\begin{align}
 d=2+\mu-\frac{2\mu\gamma(1+\mu)}{\gamma(1+\mu)+1}.\label{pt-ep-ckpzlong}
\end{align}
The same obtained from a fixed dimension RG scheme is given by
\begin{align}
 2-d=\frac{(2+\mu-d)[2\gamma(d-1)+\mu-d]}{2[2\gamma(d-1)+2+\mu-d]}.\label{pt-fd-ckpzlong}
\end{align}
As expected and similar to the quenched disordered KPZ equation studied above, the location of these lines in the $\mu-d$ plane depend explicitly on $\gamma$.  The surface is {\em logarithmically rough} on the line $\chi_h=0$. The dynamic exponent $z$ on this line will be $4-\frac{\mu[\gamma(1+\mu)-1]}{\gamma(1+\mu)+1}$ by using (\ref{pt-ep-ckpzlong}) for $\epsilon_2$ expansion case or $2+d$ by using (\ref{pt-fd-ckpzlong}) for fixed dimension RG case.

 We find that at 2D, $\epsilon_2=\mu$ in the $\epsilon_2$-expansion method, and hence from (\ref{chi-ckpz-long}) $\chi_h=\frac{\mu}{4}[-1+\frac{1}{\gamma(1+\mu)}]$. We further note that again in the fixed dimension RG method, by using $d=2$, (\ref{chi-fixed-ckpzlong}) gives $\chi_h=-\frac{\mu(2\gamma+\mu-2)}{4(2\gamma+\mu)}$. On the whole, these show that $\chi_h$ can be negative depending on values of $\gamma$ and $\mu$. This means that the surface can be smooth due to presence of long range disorder instead of logarithmically rough in pure CKPZ case.

Precisely  at dimension $d=2+\mu$, the flow equation of $g_2$ is
\begin{align}
 \frac{d\varTheta_3}{d\ell}=-\frac{1+\mu}{2+\mu}4\gamma\varTheta_3^2.\label{flow-long-dc-ckpz}
\end{align}
Equation~(\ref{flow-long-dc-ckpz}), which has the same structure as (\ref{flow-long-dc-kpz}), reveals that $\varTheta_3(\ell)$ is marginally irrelevant and it flows to zero with increasing of $\ell$; in fact $\varTheta_3=0$ is the {\em only} fixed point, which  is stable. Furthermore, $\varTheta_3(\ell)$ scales as $\frac{2+\mu}{4\gamma(1+\mu)}\frac{1}{\ell}$ in the limit of large renormalisation time length ($\ell$), same as the behaviour of $g_3(\ell)$ for large $\ell$ in Sec.~\ref{corr-kpz} above. This behaviour of coupling constant realises  the leading sacling: dynamic exponent $z=4$ and roughness exponent $\chi_h=-\mu/2$. This in turn gives breakdown of conventional dynamic scaling with 
\begin{equation}
t\sim r^4(\log(r/a_0))^{-\frac{\gamma(1+\mu)-1}{2\gamma(2+\mu)}},
\end{equation}
meaning the dynamics is logarithmically faster than the linear theory. Furthermore, the extent of ``logarithmic speeding up'' vis-a-vis the linear theory depends explicitly on $\gamma$, and also unsurprisingly on $\mu$. 
 The variance of $h$ at $d=2+\mu$ is
\begin{align}
 \langle h^2({\bf x},t)\rangle\simeq\int_{1/L}^{1/a} \frac{d^{2+\mu}{\bf q}}{q^{2}}[\log(1/q)]^{-\frac{\gamma(1+\mu)-1}{2\gamma(2+\mu)}}. \label{h2-kpzlong1}
\end{align}
The above expression says that variance does not diverge for large $L$.

So far we have studied $\mu\geq 0$ above. In principle, one may also consider the case $\mu<0$. This choice makes $\varTheta_3$ subdominant to $\varTheta_1$ at all dimensions, making the model identical to the pure CKPZ equation in the long wavelength limit. In fact, as in the case of quenched disorder KPZ equation above, the $\mu<0$ case for the CKPZ equation is statistically identical to the  pure ckpz case in the long wavelength limit. { Interestingly, the $\mu=0$ case for the quenched disordered KPZ equation is statistically identical to the pure KPZ equation in the long wavelength limit; in contrast, the $\mu=0$ case for the quenched disordered CKPZ equation belongs to a new universality class {\em different} from the pure CKPZ equation. This amply highlights the role of the conservation law in determining the universality classes in driven diffusive systems.}

\begin{figure}[!h]
\includegraphics[width=\columnwidth]{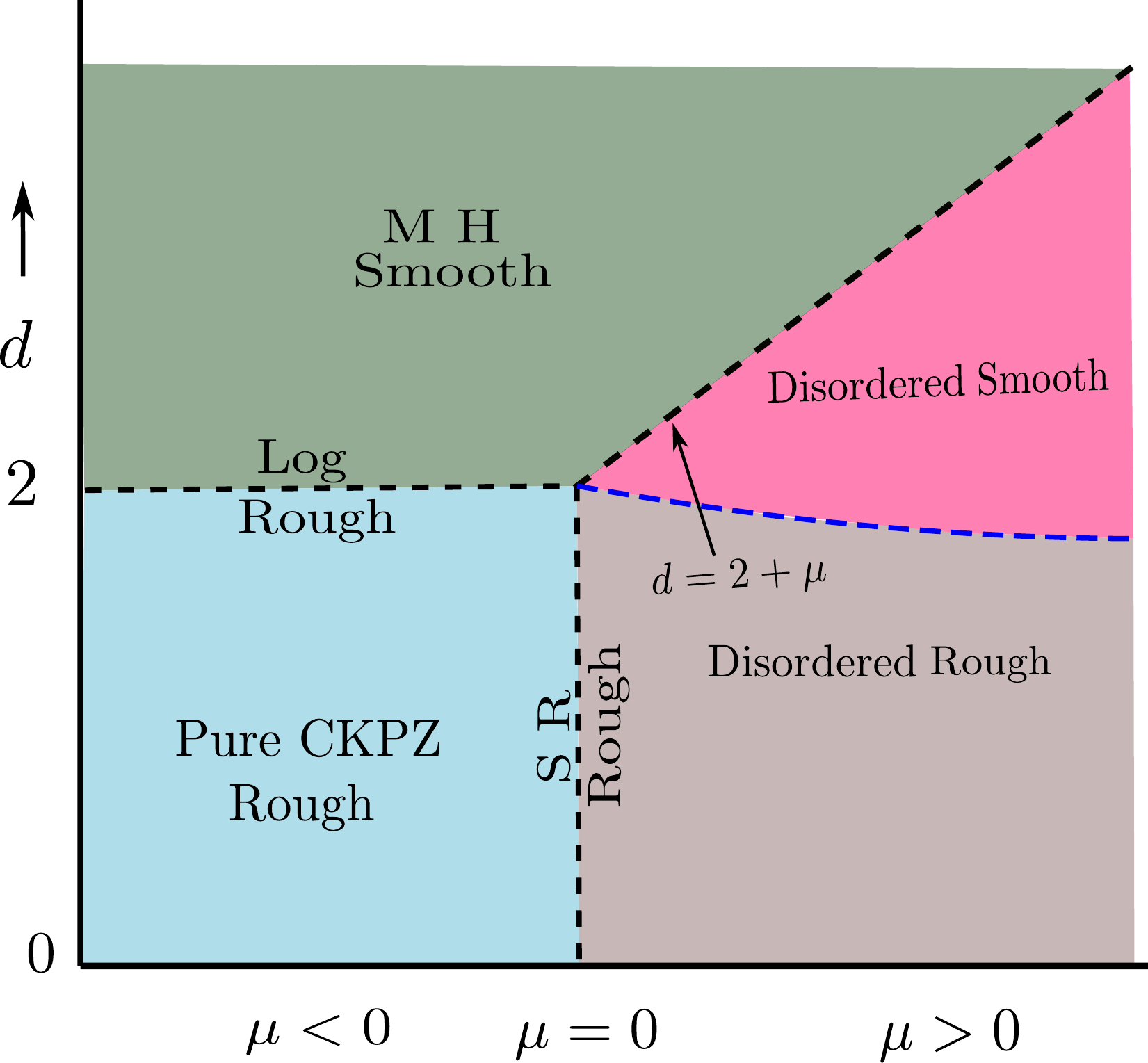}
\caption{(color online) Schematic phase diagram of the disordered CKPZ equation in the $\mu-d$ plane. The broken blue line divides disorder induced rough and smooth phases, and depends on disorder parameter $\gamma$; here this line is schematically drawn for $\gamma=1$).}\label{totalphase-ckpz}
 \end{figure}

We summarise our results on the disordered CKPZ equation in the form of  a phase diagram in Fig.~\ref{totalphase-ckpz} showing all the possible phases in $\mu-d$ plane.

We thus obtain the  scaling exponents varying continuously with $\gamma$ for both the short and long range disorder cases.

 { We summarise our results on the scaling exponents of the quenched disordered KPZ and CKPZ equations in the following  Table~\ref{tab1} and Table~\ref{tab2}.
 \linebreak
 \begin{table*}[!hbt]
 \begin{center}
 \begin{tabular}{ |m{0.91\columnwidth}|m{1.11\columnwidth}| }
 \hline
 \multicolumn{2}{|c|}{Disordered KPZ equation} \\
 \hline
 $\mu=0$: uncorrelated disorder & $\mu>0$: correlated disorder\\
 \hline
 \begin{tabular}{ p{0.17\columnwidth}|p{0.4\columnwidth}|p{0.3\columnwidth} |p{0.17\columnwidth}|p{0.53\columnwidth}|p{0.37\columnwidth}|}
 \hline
 Dimension & Exponents & Possible phases & Dimension & Exponents & Possible phases \\
 \hline
{$d<2$} & Pure KPZ result: disorder irrelevant. At $d=1$, $z=3/2$ and $\chi_h=1/2$. & Rough & $d<2+\mu$ &  $\epsilon_1$ expansion:
                `                              \begin{eqnarray}
                                               &z=2-\frac{\epsilon_1}{2\gamma}\frac{\gamma(1+\mu)-1}{(1+\mu)},\nonumber\\ 
                                               &\chi_h=\frac{\epsilon_1}{4}[1+\frac{3+\mu}{\gamma(1+\mu)}]-\frac{\mu}{2}.\nonumber
                                              \end{eqnarray}
                                              Fixed dimension RG:
                                              \begin{eqnarray}
 &z=2-\frac{(2+\mu-d)[2\gamma(d-1)+\mu-d]}{2[2\gamma(d-1)+2+\mu-d]},\label{z-fixed-longkpz}\nonumber\\
 &\chi_h=\frac{2-d}{2}-\frac{(2+\mu-d)[2\gamma(d-1)+\mu+d]}{4[2\gamma(d-1)+2+\mu-d]}.\nonumber
\end{eqnarray}
                                              
   & Disorder induced rough and smooth possible. At the transition $\chi_h=0$ with $z=2+\frac{\mu[1-\gamma(1+\mu)]}{3+\mu+\gamma(1+\mu)}$ and $z=d+\frac{d(2+\mu-d)}{2\gamma(d-1)+2+\mu-d}$ for $\epsilon_1$ expansion and  fixed dimension RG case respectively. \\
     \hline
     $d=2=d_{lc}$ & Pure KPZ result. & Rough & $d=2+\mu=d_{uc}$ & $z=2$, $\chi_h=-\mu/2$. & Smooth\\
     \hline
     $d>2$ & EW result: $z=2,\, \chi_h=\frac{2-d}{2}$, or the scaling exponents of the  pertubatively inaccessible rough. & Smooth and rough with a roughening transition & $d>2+\mu$ & EW result: $z=2,\, \chi_h=\frac{2-d}{2}$. & Smooth\\
     \hline
     \end{tabular}
 \end{tabular}
 \caption{Scaling exponents for the disordered KPZ equation as obtained  from both $\epsilon_1$-expansion and fixed dimension RG.}
 \label{tab1}
\vskip0.3in
 \begin{tabular}{ |m{0.97\columnwidth}|m{1.07\columnwidth}| }
 \hline
 \multicolumn{2}{|c|}{Disordered CKPZ equation} \\
 \hline
 
 $\mu=0$: uncorrelated disorder & $\mu>0$: correlated disorder\\
 \hline
 \begin{tabular}{ p{0.22\columnwidth}|p{0.39\columnwidth}|p{0.32\columnwidth}| p{0.2\columnwidth}|p{0.53\columnwidth}|p{0.3\columnwidth}|}
\hline
Dimension & Exponents & Possible phases & Dimension & {Exponents} & Possible phases \\
 \hline
$d=1<d_{uc}=2$ & $\epsilon_2$ expansion with $\epsilon_2=1$: 
                                \begin{eqnarray}
                                &z=4-\frac{\epsilon_2\gamma}{2\gamma+1},\nonumber\\
                                & \chi_h=\frac{\epsilon_2(1+\gamma)}{2(2\gamma+1)}.\nonumber
                                \end{eqnarray}
                            Fixed dimension RG: $z=4,\, \chi_h=1/2$.
                            & Rough & $d<2+\mu$ &  $\epsilon_2$ expansion:
                `           \begin{eqnarray}
                              &z=4-\frac{\epsilon_2}{2\gamma}\frac{\gamma(1+\mu)-1}{1+\mu},\nonumber\\
                              &\chi_h=\frac{\epsilon_2}{4}[1+\frac{1}{\gamma(1+\mu)}]-\frac{\mu}{2}. \nonumber
                            \end{eqnarray}
                        Fixed dimension RG:
                        \begin{eqnarray}
                        &z=4-\frac{(2+\mu-d)[2\gamma(d-1)+\mu-d]}{2[2\gamma(d-1)+2+\mu-d]},\nonumber\\
                        &\chi_h=\frac{2-d}{2}-\frac{(2+\mu-d)[2\gamma(d-1)+\mu-d]}{4[2\gamma(d-1)+2+\mu-d]}.\nonumber
                         \end{eqnarray}
   & Disorder induced rough and smooth possible. At the transition $\chi_h=0$ with $z=4-\frac{\mu[\gamma(1+\mu)-1]}{\gamma(1+\mu)+1}$ for $\epsilon_2$ expansion case and $z=2+d$ for fixed dimension RG case. \\
     \hline
      $d=2=d_{uc}$ & $z=4$ and $\chi_h=0$ with logarithmic roughness.  & Logarithmic rough or smooth & $d=2+\mu=d_{uc}$ & $z=4$, $\chi_h=-\mu/2$ & Smooth\\
     \hline
     $d>2$ & MH result: $z=4,\, \chi_h=\frac{2-d}{2}$. & Smooth & $d>2+\mu$ & MH result: $z=4,\, \chi_h=\frac{2-d}{2}$. & Smooth\\
     \hline
    \end{tabular}
 \end{tabular}
 \caption{Scaling exponents for the disordered CKPZ equation as obtained  from both $\epsilon_2$-expansion and fixed dimension RG.}
 \label{tab2}
 \end{center}
 \end{table*}
 }


 \section{Summary and outlook}\label{summ}

In summary, we have studied the universal scaling properties of the  KPZ and conserved KPZ equations coupled to orientational quenched disorders. We have chosen a particular form of disorder that couples with the local gradient of the height field, i.e., the disorder is sensitive to the local height nonuniformity. It is described by a zero-mean Gaussian distributed quenched vector field $\bf V$. Vector $\bf V$ can in general have both irrotational and solenoidal parts. This is reflected in the variance of $\bf V(k)$ having parts proportional to the transverse $P_{ij}({\bf k})$ and longitudinal projection operator $Q_{ij}({\bf k})$, with amplitudes $D_T$ and $D_L$, respectively. This allows us to define a dimensionless number $\gamma\equiv D_T/D_L$.  We have studied the scaling properties of the models for both spatially short and long ranged quenched disorders by using a one-loop perturbative RG scheme.

For each of the models, we have calculated the relevant scaling exponents, i.e., the roughness exponent $\chi_h$ and dynamic exponent $z$. For instance, we find that  the disordered KPZ equations with short ranged quenched disorder belongs to the well-known KPZ equation only, i.e., the disorder is irrelevant in the RG sense. Unsurprisingly, $d=2$ remains the lower critical dimension of the model, just as it is for the pure KPZ equation. For long range disorder, the model no longer belongs to the KPZ universality class; a new universality class emerges, with the pure KPZ nonlinear term being irrelevant in the RG sense. { Our result on long range orientational order induced by long range quenched disorder, when $h$ is considered as a phase, is theoretically intriguing, and should stimulate further work on disordered active XY models. }The asymptotic long wavelength scaling is controlled by the disorder nonlinearity only, and the associated scaling exponents depend explicitly on $\gamma$ and $\mu$; the latter parameter characterises the spatial scaling of the disorder variance.  In contrast, for the disordered CKPZ equation, even with short range disorder we find that   a new universality class emerges at dimension $d\leq 2$; the scaling exponents are calculated by using a one-loop expansion, which depend explicitly on $\gamma$. Further, $d=2$ is now the upper critical dimension, as in the pure CKPZ equation. For long range disorder, the pure CKPZ nonlinearity is irrelevant in the RG sense; the asymptotic long wavelength properties are controlled by the disorder nonlinearity only. This is exactly analogous to the universality in the quenched disordered KPZ with long range disorder. 
We show that as $\mu>0$ is varied, both the quenched disordered KPZ and CKPZ equations can have a transition between a rough phase and a smooth phase. Lastly, the scaling exponents, which are calculated at the one-loop order, depend explicitly on $\gamma$ and $\mu$.

Beyond the specific results, we find the surprising generic conclusion that whenever quenched disorders are  relevant (in the RG sense), the parameter $\gamma$ enters into the expressions of the scaling exponents. Since $\gamma$ can in principle vary continuously (between zero and infinity), so do the scaling exponents and hence does the universality class itself. We thus find {\em continuously varying universality}, induced by quenched disorder, in the two models that we considered. Furthermore, the existence of continuously varying universality appears to be quite robust: it exists with (CKPZ) or without (KPZ) a conservation law, so long as the quenched disorder itself remains relevant in the RG sense.  We can relate this with the symmetry of the disorder distribution: $\gamma$ is the parameter that determines the relative strengths of   the transverse and longitudinal parts of the disorder variance. Thus varying $\gamma$ effectively means variation of the local alignment of $\bf V({\bf q})$ with respect to $\bf q$. This continuous variation of the universality class is in fact more robust than our results perhaps imply. Since the variance of a quenched disordered vector field can generally be a combination of transverse and longitudinal parts, a dimensionless parameter like $\gamma$ automatically arises, which remains a free parameter, continuously varying universal properties are generically expected so long as the quenched disorder remains relevant in the RG sense. These results may be tested by designing and simulating appropriate agent-based models coupled to a vector quenched disorder field with specified distributions. Do these  results apply to annealed disorders as well? As shown in Appendix \ref{ann}, the CKPZ equations, when driven by vector fields with linear dynamics show this behaviour. Preliminary results on the KPZ equation driven by an annealed disorder field also shows similar behaviour (not shown here). This however comes with a caveat. An annealed field has a dynamics, which is not necessarily linear in general. We have assumed a linear dynamics for the annealed field only for simplicity. In general nonlinear terms should be present, which may or may not be relevant in the RG sense. If it is relevant, then there is a possibility the dynamics of the annealed disorder field is renormalized, and that a parameter equivalent to $\gamma$ may acquire specific RG fixed point value(s), instead of being a free, continuously varying parameter. Thus it is possible that the resulting universality class of the height field (or in general a dynamical field driven by the annealed disorder) will {\em not} vary continuously. However, in the event the nonlinear effects in the dynamics of the annealed field are irrelevant, we expect continuously varying universality class to follow, as we have already found in the CKPZ equation with annealed disorder. We therefore conclude that, the existence of continuously varying universality classes with a vector quench disorder field is likely to be more generic than its annealed counterparts. Lastly, we caution the reader that the continuously varying universality class observed with quenched disorder here does not hold for {\em any} kind of  quenched disorder, even if the disorder is relevant in the RG sense. This is because, in order for this result to hold, there must a free dimensionless parameter like $\gamma$ in the disorder distribution. Not all quench disorder distributions may allow this. For instance, there is no scope of this in the recent studies on quenched columnar disordered KPZ equation~\cite{astik-prr,astik-pre}, since the disorder there is a scalar.

 { Our results may be verified by direct numerical simulations of the stochastically driven equations of motion, which may be conveniently done by using pseudospectral methods~\cite{ab-burgers2}. It would be interesting to develop equivalent quenched disordered agent-based lattice-gas models, in terms of a height field or a phase field, which may be studied by using Monte-Carlo simulations and explore and validate the RG results above. We hope our studies will provide new impetus in these directions.} 

Apart from their obvious theoretical interests, our results have experimental implications as well. For instance, a possible physical origin of the (vector) quenched disorder is the underlying disordered substrates. The substrates, if not very carefully prepared, may have different patches, characterised by different values of the parameter $\gamma$.  For sufficiently large patches, each such patch should be characterised by scaling exponents parametrised by $\gamma$. If the experimental measurements are done on a region  that includes may such patches having different $\gamma$, the measured correlation functions are not likely to show any clean scaling, but should rather display 
a broadening of the measured values of the scaling exponents, or a kind of smeared scaling behaviour. We hope our studies here will provide further impetus to research along these lines in the future.


\section{Acknowledgement}

A.B. thanks the SERB, DST (India) for partial financial support through the MATRICS scheme [file no.: MTR/2020/000406].

 \appendix

 \section{Generating functional}\label{generating}

 The generating functional of the disordered KPZ or CKPZ equation has the form~\cite{martin,zinn,bausch,dedominicis} 
 \begin{align}
  \mathbb{Z}=\int_{{\bf x},t} \mathcal{D}\hat{h} \mathcal{D}h \mathcal{D}V  e^{-\mathnormal{S}[\hat{h}, h, V]}. \label{gen-fun}
 \end{align}
 Here $\mathnormal{S}$ is the {\em action functional} and it contains two parts: (i) harmonic ($\mathnormal{S_H} $) (ii) anharmonic ($\mathnormal S_A $). Field $\hat{h}({\bf x},t)$ is the dynamic conjugate field of $h ({\bf x},t)$~\cite{bausch}.

 We here write the down the explicit forms of the action functionals for the disordered KPZ and CKPZ equations.
 
 \subsection{Disordered KPZ equation}\label{kpz-action}
  The  action functional is constructed by using  Eq. (\ref{heq-kpz}) and averaging over the Gaussian-distributed noise $\xi_h$. We get
 
 \begin{align}
  \mathnormal{S^\text{KPZ}_H}=&\int_{{\bf x},t}\left[ -D_1\hat{h}\hat{h} + \hat{h}\left(\partial_t-\nu_1\nabla^2 \right)h \right] \nonumber\\
 & +\int_{\bf x,x'} \frac{V_{i_x} V_{j_{x'}}}{4[D_T P_{ij} + D_L Q_{ij}]|x-x'|^{-\alpha}},\label{ac-lin-kpz}\\
 \mathnormal {S^\text{KPZ}_A}=& \int_{{\bf x},t} -\hat{h}\left[\frac{\lambda_1}{2} (\boldsymbol \nabla h)^2 + \kappa_1 ({\bf V}.\boldsymbol\nabla h) \right]\label{ac-nonlin-kpz},
 \end{align}
 where $S^\text{KPZ}_H$ and $S^\text{KPZ}_A$ are, respectively, the harmonic and anhamornic parts of the action functional.

 \subsection{Disordered CKPZ equation}\label{ckpz-action}
 The action functional is constructed by using (\ref{heq-ckpz}) and then averaging over the Gaussian-distributed noise $\eta_h$. We get
 \begin{align}
  \mathnormal{S^\text{CKPZ}_H}=&\int_{{\bf x},t}\left[ -D_2(-\nabla^2)\hat{h}\hat{h} + \hat{h}\left(\partial_t+\nu_2\nabla^4 \right)h \right] \nonumber\\
 & +\int_{\bf x,x'} \frac{V_{i_x} V_{j_{x'}}}{4[D_T P_{ij} + D_L Q_{ij}]|x-x'|^{-\alpha}},\label{ac-lin-ckpz}\\
 \mathnormal {S^\text{CKPZ}_A}=& \int_{{\bf x},t} -\hat{h}\left[\frac{\lambda_2}{2} \nabla^2(\boldsymbol \nabla h)^2 + \kappa_2 \nabla^2({\bf V}.\boldsymbol\nabla h) \right]\label{ac-nonlin-ckpz},
 \end{align}
 where $S^\text{CKPZ}_H$ and $S^\text{CKPZ}_H$ are the harmonic and anharmonic parts of the action functional.

\section{Harmonic theory}\label{linear}

In this Section, we consider the harmonic parts of the action functionals, which correspond to the linear parts of the underlying stochastically driven equations of motion.

\subsection{Disordered KPZ equation}\label{kpz-lin}
The propagator and correlator of $h$ in the Fourier space can be obtained from  action (\ref{ac-lin-kpz}). These are 
\begin{subequations}
 \begin{align}
  &\langle \hat{h}_{-{\bf k},-\omega} h_{{\bf k},\omega} \rangle =\frac{1}{-i\omega+\nu_1 k^2},\\
  &\langle |h_{{\bf q},\omega}|^2 \rangle =\frac{2D_1 }{\omega^2+\nu_1^2k^4}.
 \end{align}
\end{subequations}
We thus get dynamic exponent $z=2$ at all dimensions. The equal-time correlator in real space   
\begin{align}
 C(r)\equiv\langle [h({\bf x},t)- h({\bf x'},t)]^2\rangle=2 \int \frac{d^d {\bf k}}{(2\pi)^d} \frac{D_1 [1- e^{i{\bf k\cdot [x-x']}}]}{\nu_1 k^2}, \label{lin-corr-kpz}
\end{align}
for large $r\equiv |{\bf x-x'}|$.
Following  (\ref{exponent}), we conclude that $\chi_h=\frac{2-d}{2}$. Therefore, $\chi_h=1/2$ at 1D and $\chi_h=0$ at 2D in the Gaussian theory, or in the linearized equation (\ref{heq-kpz}).

\subsection{Disordered CKPZ equation}\label{ckpz-lin}
The bare propagator and correlator of $h$ in Fourier space can be obtained from the harmonic part of action~(\ref{ac-lin-ckpz}). These are
\begin{subequations}
 \begin{align}
  &\langle \hat{h}_{-{\bf k},-\omega} h_{{\bf k},\omega} \rangle =\frac{1}{-i\omega+\nu_2 k^4}.\\
  &\langle |h_{{\bf q},\omega}|^2 \rangle =\frac{2D_2 k^2}{\omega^2+\nu_2^2k^8}.
 \end{align}
\end{subequations}
Here $z=4$, and the equal-time correlator $C(r)$ for this case in real space can be defined similarly as eq. (\ref{lin-corr-kpz}) which gives $\chi_h=\frac{2-d}{2}$. It  means $\chi_h=1/2$ at 1D and $\chi_h=0$ at 2D in the Gaussian theory, or in the linearised Eq.~(\ref{heq-ckpz}).

 \section{Dynamic renormalisation group (RG) method}\label{DRG}
 

 We have discussed above about the method of dynamic RG, so, here we present the intermediate details such as loop diagrams, rescaling, fluctuation corrections.
\subsection{One-loop Feynman diagrams}\label{loop}

\begin{figure}
 \includegraphics[width=\columnwidth]{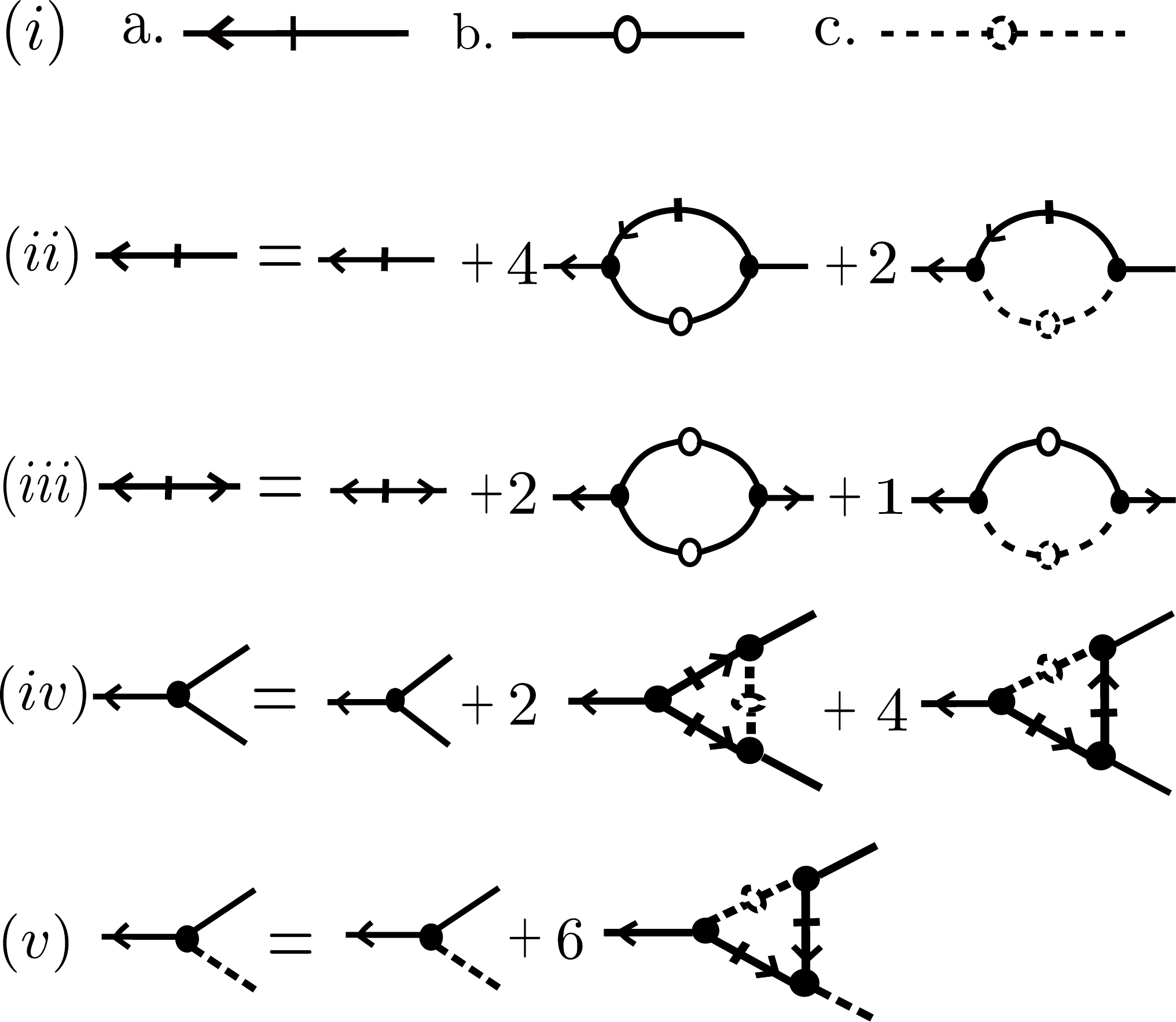}
 \caption{$(i)$ Bare two point functions: a. propagator, b. correlator, c. disorder correlator.  Figures $(ii)-(v)$ give the Feynman diagrams for the one-loop corrections in the theory; { the { one-loop}  diagrams  { have the same structure for} { in both the disordered KPZ and CKPZ equations, though their values differ (see text)} }. Figure $(ii)$ give one-loop corrections to the diffusivity ($\nu_1$ or $\nu_2$), figure $(iii)$ are for the noise strength ($D_1$ or $D_2$), figure $(iv)$ are for the pure nonlinear terms ($\lambda_1$ or $\lambda_2$), and figure $(v)$ correspond to the disordered nonlinear terms ($\kappa_1$ or $\kappa_2$). The number written before a Feynman diagram is the symmetry factor of the corresponding diagram.}\label{feynmann}
\end{figure}

We represent the bare two point functions, the anharmonic terms or the ``vertices'' in the action functional and the fluctuation-corrections of the model parameters by the Feynman diagrams, as shown in Fig. \ref{feynmann}. These diagrams are for both the disordered KPZ and CKPZ equations. However, their expressions differ depending upon the model in question.

\subsubsection{Disordered KPZ equation}
The fluctuation-corrections of the model parameters of the disordered KPZ equation~(\ref{heq-kpz}) corresponding to the one-loop diagrams in Fig.~\ref{feynmann} are given below.

\begin{widetext}
\begin{subequations}
 
 \begin{eqnarray}
  &&{\rm Fig.~}\ref{feynmann}(ii):~~ \nu_1^<=\nu_1\left[ 1+\frac{2-d}{4d}\frac{\lambda_1^2D_1}{\nu_1^3}\int_{\varLambda/b}^{\varLambda}\frac{d^d{\bf q}}{(2\pi)^d)}\frac{1}{q^2}+\left\{\frac{d-1}{d}\frac{2\kappa_1^2D_T}{\nu_1^2}+\frac{\mu-d}{d}\frac{\kappa_1^2D_L}{\nu_1^2} \right\}\int_{\varLambda/b}^{\varLambda}\frac{d^d{\bf q}}{(2\pi)^d)}\frac{1}{q^{2+\mu}}\right].\\
  &&{\rm Fig.~}\ref{feynmann}(iii):~~ D_1^<=D_1\left[1 +\frac{\lambda_1^2 D_1}{4\nu_1^3}\int_{\varLambda/b}^{\varLambda}\frac{d^d{\bf q}}{(2\pi)^d)}\frac{1}{q^2} + \frac{2\kappa_1^2D_L}{\nu_1^2}  \int_{\varLambda/b}^{\varLambda}\frac{d^d{\bf q}}{(2\pi)^d)}\frac{1}{q^{2+\mu}}\right] .\\
  &&{\rm Fig.~}\ref{feynmann}(iv):~~ \lambda_1^<=\lambda_1\left[1+ \left\{ \frac{d-1}{d}\frac{2\kappa_1^2D_T}{\nu_1^2}-\frac{2}{d}\frac{\kappa_1^2D_L}{\nu_1^2} \right\}\int_{\varLambda/b}^{\varLambda}\frac{d^d{\bf q}}{(2\pi)^d)}\frac{1}{q^{2+\mu}} \right].\\
  &&{\rm Fig.~}\ref{feynmann}(v):~~ \kappa_1^<=\kappa_1\left[1-\frac{2}{d}\frac{\kappa_1^2D_L}{\nu_1^2} \int_{\varLambda/b}^{\varLambda}\frac{d^d{\bf q}}{(2\pi)^d)}\frac{1}{q^{2+\mu}} \right].
 \end{eqnarray}\label{parameter-correction-kpz}


\end{subequations}
 \end{widetext}

\subsubsection{Disordered CKPZ equation}
The fluctuation-corrections of the parameters in the disordered CKPZ equation~(\ref{heq-ckpz}) corresponding to the one-loop Feynman diagrams in Fig.~\ref{feynmann} are given below.
\begin{widetext}
\begin{subequations}
 \begin{eqnarray}
  &&{\rm Fig.~}\ref{feynmann}(ii):~~ \nu_2^<=\nu_2\left[ 1+\frac{4-d}{4d}\frac{\lambda_2^2D_2}{\nu_2^3}\int_{\varLambda/b}^{\varLambda}\frac{d^d{\bf q}}{(2\pi)^d)}\frac{1}{q^2}+\left\{\frac{d-1}{d}\frac{2\kappa_2^2D_T}{\nu_2^2}+\frac{\mu-d}{d}\frac{\kappa_2^2D_L}{\nu_2^2} \right\}\int_{\varLambda/b}^{\varLambda}\frac{d^d{\bf q}}{(2\pi)^d)}\frac{1}{q^{2+\mu}}\right].\\
  &&{\rm Fig.~}\ref{feynmann}(iii):~~ D_2^<=D_2 + {\rm subleading~ corrections}.\\
  &&{\rm Fig.~}\ref{feynmann}(iv):~~ \lambda_2^<=\lambda_2\left[1+ \left\{ \frac{d-1}{d}\frac{2\kappa_2^2D_T}{\nu_2^2}-\frac{2}{d}\frac{\kappa_2^2D_L}{\nu_2^2} \right\}\int_{\varLambda/b}^{\varLambda}\frac{d^d{\bf q}}{(2\pi)^d)}\frac{1}{q^{2+\mu}} \right].\\
  &&{\rm Fig.~}\ref{feynmann}(v):~~ \kappa_2^<=\kappa_2\left[1-\frac{2}{d}\frac{\kappa_2^2D_L}{\nu_2^2} \int_{\varLambda/b}^{\varLambda}\frac{d^d{\bf q}}{(2\pi)^d)}\frac{1}{q^{2+\mu}} \right].
 \end{eqnarray}\label{parameter-correction-ckpz}
\end{subequations}
 \end{widetext}

 \subsection{Rescaling}\label{rescale}
We rescale a Fourier wavevector and frequency as ${\bf k}\rightarrow b {\bf k}$ and $\omega\rightarrow b^z \omega$, equivalently in the real space ${\bf x}\rightarrow{\bf x}/b,\, t\rightarrow t/b^z$, where $z$ is the dynamic exponent. The fields are accordingly rescaled as
\begin{align}
 h({\bf k},\omega)=\xi h(b{\bf k},b^z\omega);\, \hat{h}({\bf k},\omega)=\hat{\xi} \hat{h}(b{\bf k},b^z\omega).
\end{align}
Furthermore, $V_i({\bf k},\omega)=b^{z+\frac d2 +\frac \mu2} V_i(b{\bf k},b^z\omega)$, as obtained from Eq.~(\ref{v-var}).

\subsubsection{Scaling of the parameters in the disordered KPZ equation}
We set $\xi \hat{\xi}= b^{d+2z}$ by demanding the term in {\em action} (\ref{ac-lin-kpz}) $\int d^d{\bf k} d\omega \hat{h} h\omega$ does not scale under the rescaling of wavevectors and frequencies. Accordingly the model parameters scale as given below.
\begin{align}
 &\nu_1'=\nu_1^<b^{z-2},\, D'_1=D_1^< b^{z-d-2\chi_h},\nonumber\\
 &\lambda_1'=\lambda_1^< b^{z-2+\chi_h},\,  \kappa'_1=\kappa_1 b^{z-1+\frac \mu2-\frac d2},\nonumber \\
 &D_L'=D_L^<=D_L,\,\,D_T'=D_T^< = D_T.
\end{align}

\subsubsection{Scaling of the parameters in the disordered CKPZ equation}
We set $\xi \hat{\xi}= b^{d+2z}$ by demanding the term in the action functional (\ref{ac-lin-ckpz}) $\int d^d{\bf k} d\omega \hat{h} h\omega$ does not scale under the rescaling of wavevectors and frequencies. Accordingly the model parameters scale as given below.
\begin{align}
 &\nu_2'=\nu_2^<b^{z-4},\, D'_2=D_2^< b^{z-d-2-2\chi_h},\nonumber\\
 &\lambda_2'=\lambda_2^< b^{z-4+\chi_h},\,  \kappa_2'=\kappa_2 b^{z-3+\frac \mu2-\frac d2}\nonumber \\
 &D_L'=D_L^<=D_L,\,\,D_T'=D_T^< = D_T.
\end{align}

\section{Annealed disorder}\label{ann}

We now briefly discuss what happens when the disorder is ``annealed'' instead of ``quenched''. This means the disorder is no longer frozen in time. At the simplest level, we model it by a time-dependent fluctuating vector field ${\bf U}({\bf x},t)$ that has zero mean and is assumed to be Gaussian distributed. If $\bf U$ is interpreted as a velocity field, then this is reminiscent of the well-known passive scalar turbulence problem~\cite{Kraichnan,Adzhemyan,Tirthankar,sudip-passive}. The Eq. of motion of $\bf U$ is 
\begin{align}
\frac{\partial \bf U}{\partial t}=b\nabla^2 {\bf U}+{\bf f}, \label{U-eom}
\end{align}
where the noise ${\bf f} $ is zero-mean, Gaussian-distributed with a variance 
\begin{align}
 &\langle f_i({\bf k},\omega) f_j({\bf k'},\omega') \rangle \nonumber\\
 &= [2a_1 P_{ij}+2a_2 Q_{ij}]k^{2-\mu}\delta^d({\bf k-k'})\delta(t-t').\label{f-var}
\end{align}
Here, $\mu=d-\alpha$, as in the main text. Then the  velocity autocorrelation  function can be calculated exactly. It in Fourier space is 
 \begin{align}
  &\langle U_i({\bf k},\omega) U_j({\bf k'},\omega') \rangle= \frac{2a_1 P_{ij}+2a_2 Q_{ij}}{\omega^2+b^2k^4}~ k^{2-\mu}\nonumber\\
 &~~~~~~~~~~~~~~~~~~~~\times\delta^d({\bf k+k'}) \delta(\omega+\omega'),\label{v-var-ann}
 \end{align}
 where,  $a_1/b=2\tilde D_T$ and $a_2/b=2\tilde D_L$. We are interested in the large $b$ limit, with $\tilde D_L$ and $\tilde D_T$ being finite. In this limit, the annealed disorder autocorrelation in real space reduces to
\begin{align}
 &\langle U_i({\bf x},t) U_j({\bf x'},t') \rangle\nonumber\\
 &= \left[2\tilde D_T P_{ij}+2\tilde D_L Q_{ij}\right]|{\bf x-x'}|^{-\alpha}\delta(t-t'). \label{veq-ann}
\end{align}

 
 We briefly discuss the scaling properties of the disordered CKPZ equation with annealed disorder of the type defined above. The one-loop Feynman diagrams are identical to their quenched disorder counterparts. We focus on the one-loop diagrams which vanish in the absence of the annealed disorder.
 
 Let us consider the last diagram in Fig.~\ref{feynmann}(ii) that originates from the disorder coupling $\kappa_2$ and   corrects $\nu_2$:
 \begin{align}
  \int_{\bf q} \frac{({\bf k-q})_i({\bf k-q})^2q^{2-\mu}(2a_1P_{ij}+2a_2Q_{ij})_q}{bq^2[bq^2+\nu_2({\bf k-q})^4]},\label{nu2-annealed}
 \end{align}
 where {\bf k} is an external wavevector and {\bf q} is an internal wavevector of the one-loop diagram.
Noise strength $D_2$ receives  no relevant or diverging corrections. 

In Fig. {\ref{feynmann}(iv) the triangle diagrams that originate from the disorder coupling $\kappa$  correct $\lambda_2$. They are
\begin{align}
 &{\rm first\;\;triangle\;\;diagram}\sim \int_{\bf q} q^6q^{2-\mu}[2a_1P_{im}+2a_2Q_{im}]\nonumber\\
 &[\frac{1}{bq^2[\nu_2^2q^8-b^2q^4]}-\frac{1}{\nu_2q^4[b^2q^4-\nu_2q^8]}].\label{lambda2-1-annealed}\\
 & {\rm second\;\;triangle\;\;diagram}\sim \nonumber\\
 &~~~~~~\int_{\bf q} \frac{q_mq_jq^4q^{2-\mu}[2a_1P_{im}+2a_2Q_{im}]}{bq^2[bq^2+\nu_2q^4]^2}.\label{lambda2-2-annealed}
\end{align}
Similarly, the last diagram
 in Fig. {\ref{feynmann}(v), coming from the disorder vertex,  corrects $\kappa_2$.
 \begin{align}
  \int_{\bf q} \frac{q_mq_jq^4q^{2-\mu}[2a_1P_{im}+2a_2Q_{im}]}{bq^2[bq^2+\nu_2q^4]^2}.\label{kappa2-annealed}
 \end{align}
All the expressions of different one-loop diagrams given in (\ref{nu2-annealed}),(\ref{lambda2-1-annealed}),(\ref{lambda2-2-annealed}),(\ref{kappa2-annealed}) are infrared divergent if $b\ll\nu_2q^2$ is satisfied. In fact, these corrections are same as in (\ref{parameter-correction-ckpz}).  However, these diagrams give subleading corrections to the parameters in the opposite limit $b\gg\nu_2q^2$.
 
 As the high wavevector modes are being eliminated starting  from $k=\varLambda$, the upper waveveector curoff, there can be two situations: either $\nu_2k^2$ dominates over $b$ over a substantial range of (i.e., intermediate values of) wavevectors just below $\varLambda$, or $b$ dominates over $\nu_2 k^2$ in the same wavevector region. 
 Noting that with $\nu_2$ gaining positive corrections but $b$ gets none, there can be two distinct possibilities.  
 
 (I) It is possible that as one eliminates the higher wavevector modes, {\em before} $b$ starts dominating, long enough RG time is spent, and the annealed-disordered CKPZ equation is {\em renormalised}. In this case,  The corrections of parameters and corresponding flow equations are same as equations of (\ref{parameters flow ckpz}). Therefore the result is identical to quenched disordered CKPZ equation. In this case, for even lower wavevector regimes with $b\gg \nu_2k^2$, there are no new relevant fluctuation corrections to the model parameters. Hence, the same universality as the quenched disordered CKPZ equation ensues, with a continuously varying parameter $\tilde\gamma\equiv \tilde D_T/\tilde D_L$ parametrising the universality class.
 
 (II) In the opposite case, not enough RG time is available to make $\nu_2$ substantially renormalise, making $b$ dominate over $\nu_2 k^2$ for low enough $k$. In this case, further more elimination from the disorder-dependent diagrams do not produce any relevant corrections to the parameters. However, the fluctuation corrections from the pure CKPZ nonlinear terms still survive, ultimately giving the scaling exponents of the pure CKPZ equation. In this case, naturally, there are no $\tilde\gamma$-dependence of the scaling exponents.

 \bibliography{ckpzdisorder.bib}

\end{document}